\documentclass[10pt,twocolumn,letterpaper]{article}

\usepackage{cvpr}
\usepackage{times}
\usepackage{epsfig}
\usepackage{graphicx}
\usepackage{amsmath}
\usepackage{amssymb}

\usepackage{enumitem}
\usepackage[nocompress]{cite}
\usepackage{verbatim}
\usepackage{multirow}
\usepackage{subfigure}
\usepackage{mathrsfs}
\usepackage{booktabs}  
\usepackage{threeparttable}
\usepackage{setspace}

\usepackage{colortbl}
\usepackage{color}
\definecolor{myBlue}{rgb}{0.6,0.6,1}
\definecolor{myRed}{rgb}{1,0.424,0.569}
\definecolor{myGreen}{rgb}{0.68,0.812,0}
\usepackage{rotating}
\usepackage{authblk}
\usepackage[misc]{ifsym}
\usepackage{bbding}

\usepackage{array}
\newcommand{\PreserveBackslash}[1]{\let\temp=\\#1\let\\=\temp}
\newcolumntype{C}[1]{>{\PreserveBackslash\centering}p{#1}}
\newcolumntype{R}[1]{>{\PreserveBackslash\raggedleft}p{#1}}
\newcolumntype{L}[1]{>{\PreserveBackslash\raggedright}p{#1}}

\makeatletter
  \newcommand\figcaption{\def\@captype{figure}\caption}
  \newcommand\tabcaption{\def\@captype{table}\caption}
\makeatother

\usepackage{xcolor,pifont}
\newcommand*\colourcheck[1]{\expandafter\newcommand\csname #1check\endcsname{\textcolor{#1}{\ding{52}}}}
\colourcheck{blue}
\colourcheck{green}
\colourcheck{red}
\usepackage{pifont}

\newcommand{\tabincell}[2]{\begin{tabular}{@{}#1@{}}#2\end{tabular}}  

\usepackage{algorithm}
\usepackage{algpseudocode}

\usepackage{url}

\usepackage[pagebackref=true,breaklinks=true,letterpaper=true,colorlinks,bookmarks=false]{hyperref}

\cvprfinalcopy

\def\cvprPaperID{809}

\ifcvprfinal\fi
\begin{document}
\definecolor{light_gray}{gray}{0.9}

\title{Camera Trace Erasing\vspace{-3ex}}
\author[$1$]{Chang Chen\thanks{This work was partially done when Chang Chen was a visiting scholar at Michigan State University.}\vspace{-1.2ex}} 
\author[$1$]{Zhiwei Xiong \Envelope \ }
\author[$2$]{Xiaoming Liu	}
\author[$1$]{Feng Wu}
\affil[ ]{$^{1}$ University of Science and Technology of China \hspace{0.9cm} $^{2}$ Michigan State University\vspace{+0.7ex}}
\affil[ ]{\small{\nolinkurl{changc@mail.ustc.edu.cn,} \ \nolinkurl{{zwxiong, fengwu}@ustc.edu.cn,} \ \nolinkurl{liuxm@cse.msu.edu} } \vspace{-1ex}}
\maketitle

\begin{abstract}
\vspace{-0.13cm}
Camera trace is a unique noise produced in digital imaging process. Most existing forensic methods analyze camera trace to identify image origins.
In this paper, we address a new low-level vision problem, camera trace erasing, to reveal the weakness of trace-based forensic methods.
A comprehensive investigation on existing anti-forensic methods reveals that it is non-trivial to effectively erase camera trace while avoiding the destruction of content signal.
To reconcile these two demands, we propose Siamese Trace Erasing (SiamTE), in which a novel hybrid loss is designed on the basis of Siamese architecture for network training.
Specifically, we propose embedded similarity, truncated fidelity, and cross identity to form the hybrid loss.
Compared with existing anti-forensic methods, SiamTE has a clear advantage for camera trace erasing, which is demonstrated in three representative tasks.
Code and dataset are available at \url{https://github.com/ngchc/CameraTE}.
\end{abstract}

\vspace{-0.35cm}
\section{Introduction}
\vspace{-0.05cm}
\label{sec:introduction}
Noise is inevitable in digital imaging process. Camera trace is such a kind of noise that is unique to each type of imaging device.
Specifically, camera trace is produced by the different response characteristics of camera sensor to light \cite{PRNU}, and then manipulated by the in-camera processing pipeline \cite{CC}.
Therefore, camera trace implicitly encodes information of camera type into the imaging results in a form of noise.
Based on the analysis on camera trace, researches have proposed a variety of methods for image forensic tasks, in terms of origin identification \cite{CamFP, ResNet-CMID}, tampering detection \cite{Chen-SPL2015, Zhou-CVPR2018, zhou2017two}, and forgery localization \cite{Lyu-IJCV2014, Noiseprint, wu2019mantra}, to name a few.
These methods play an important role in the steady development of image-based social networks \cite{zheng2019survey}.

\begin{figure}[t]
\begin{center}
\begin{minipage}{0.49\linewidth}
\begin{minipage}{0.5\linewidth}
  \centerline{\includegraphics[width=1\linewidth]{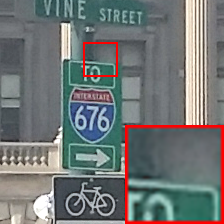}}
  \fontsize{8}{9.6}\selectfont
  \vspace{-0.05cm}
  \centerline{(a) Original}
\end{minipage}
\begin{minipage}{0.465\linewidth}
  \centerline{\includegraphics[width=1\linewidth]{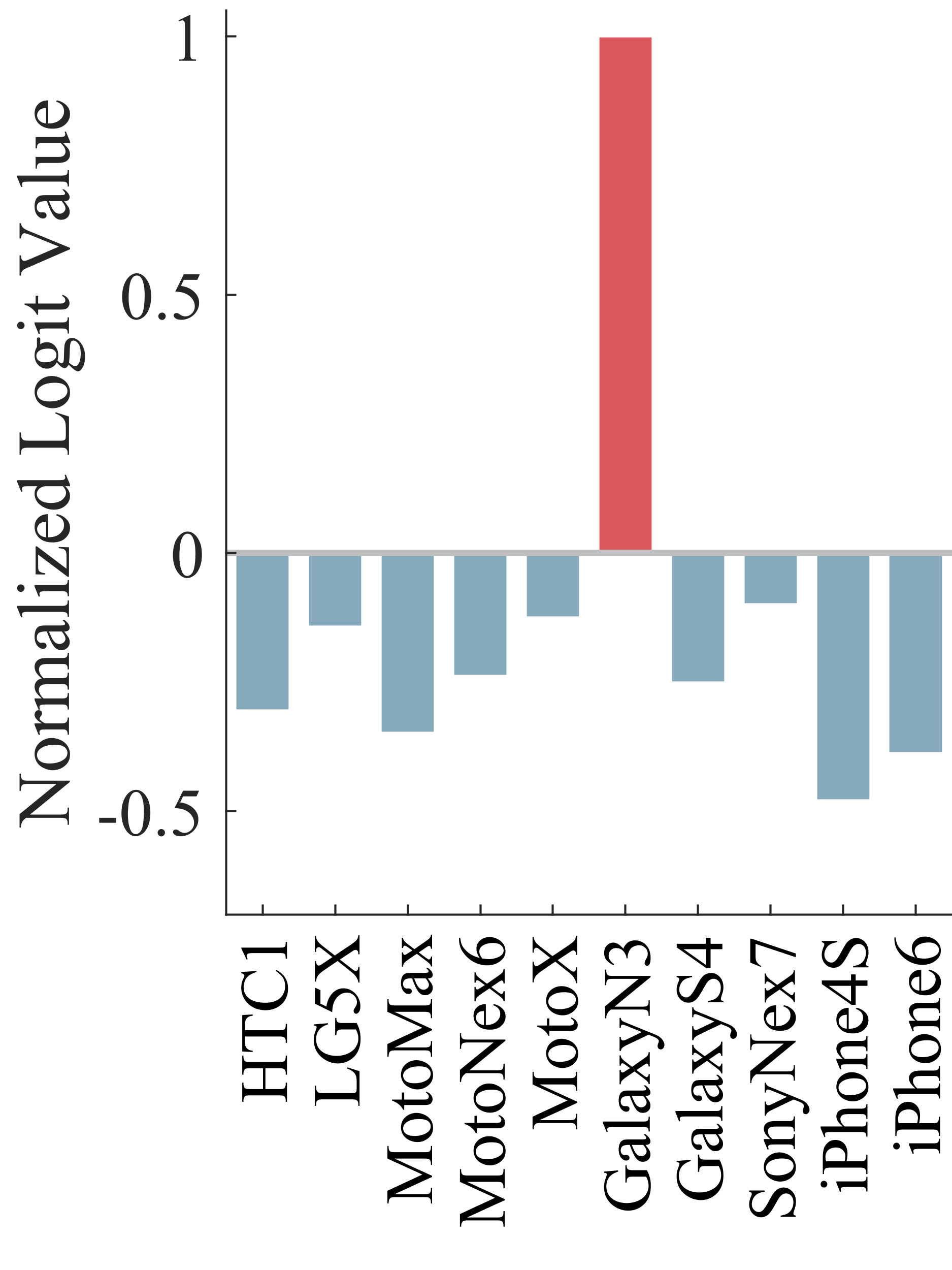}}
\end{minipage}
\end{minipage}
\hspace{-0.1cm}
\begin{minipage}{0.49\linewidth}
\begin{minipage}{0.5\linewidth}
  \centerline{\includegraphics[width=1\linewidth]{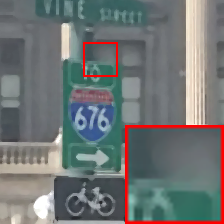}}
  \fontsize{8}{9.6}\selectfont
  \vspace{-0.05cm}
  \centerline{(b) Filtering}
\end{minipage}
\begin{minipage}{0.465\linewidth}
  \centerline{\includegraphics[width=1\linewidth]{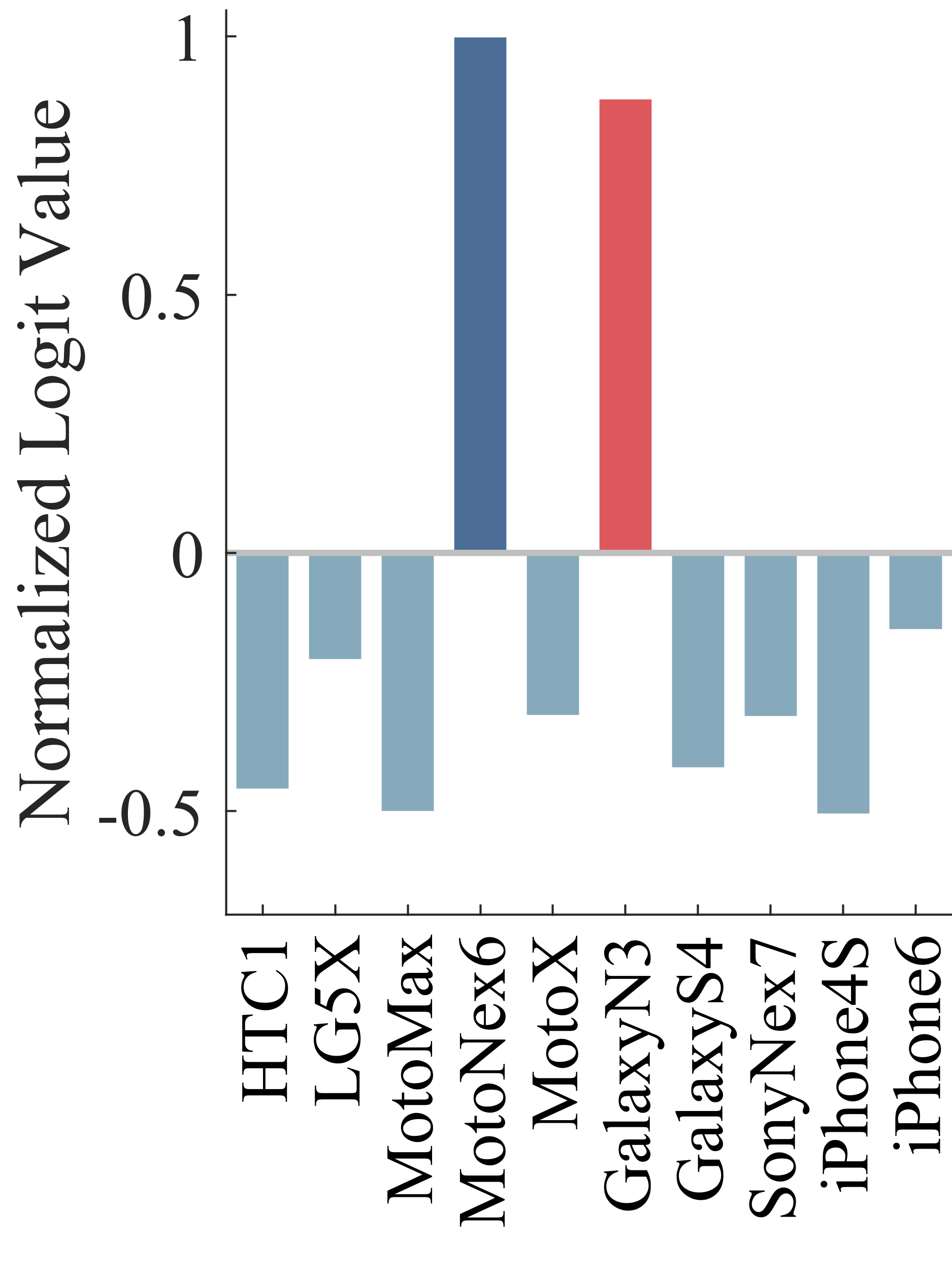}}
\end{minipage}
\end{minipage}
\vfill
\vspace{-0.05cm}
\begin{minipage}{0.49\linewidth}
\begin{minipage}{0.5\linewidth}
  \centerline{\includegraphics[width=1\linewidth]{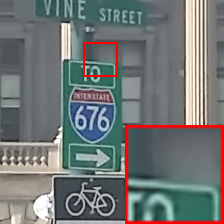}}
  \fontsize{8}{9.6}\selectfont
  \vspace{-0.05cm}
  \centerline{(c) Denoising}
\end{minipage}
\begin{minipage}{0.465\linewidth}
  \centerline{\includegraphics[width=1\linewidth]{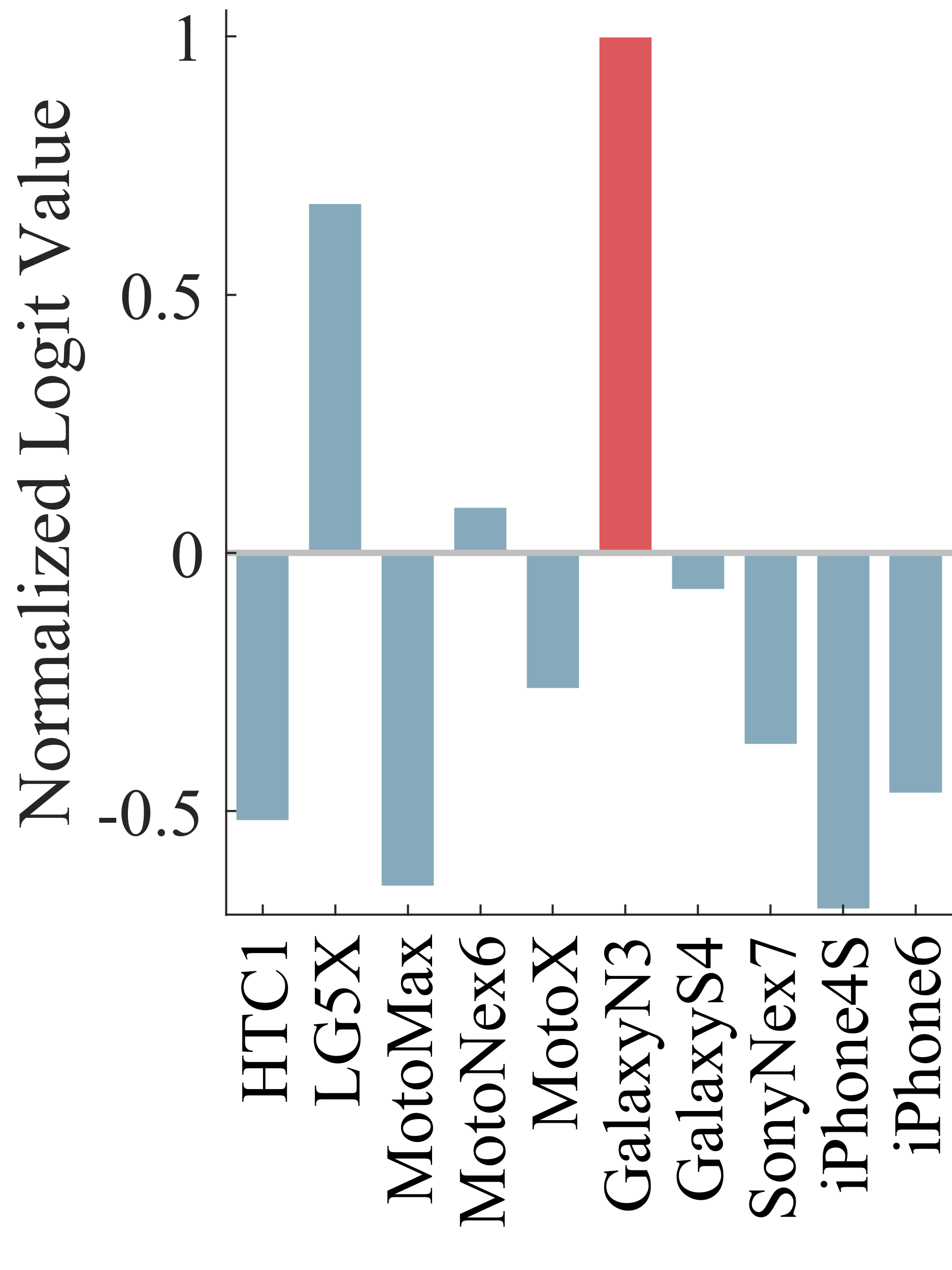}}
\end{minipage}
\end{minipage}
\hspace{-0.1cm}
\begin{minipage}{0.49\linewidth}
\begin{minipage}{0.5\linewidth}
  \centerline{\includegraphics[width=1\linewidth]{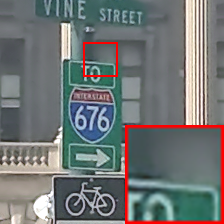}}
  \fontsize{8}{9.6}\selectfont
  \vspace{-0.05cm}
  \centerline{(d) Ours}
\end{minipage}
\begin{minipage}{0.465\linewidth}
  \centerline{\includegraphics[width=1\linewidth]{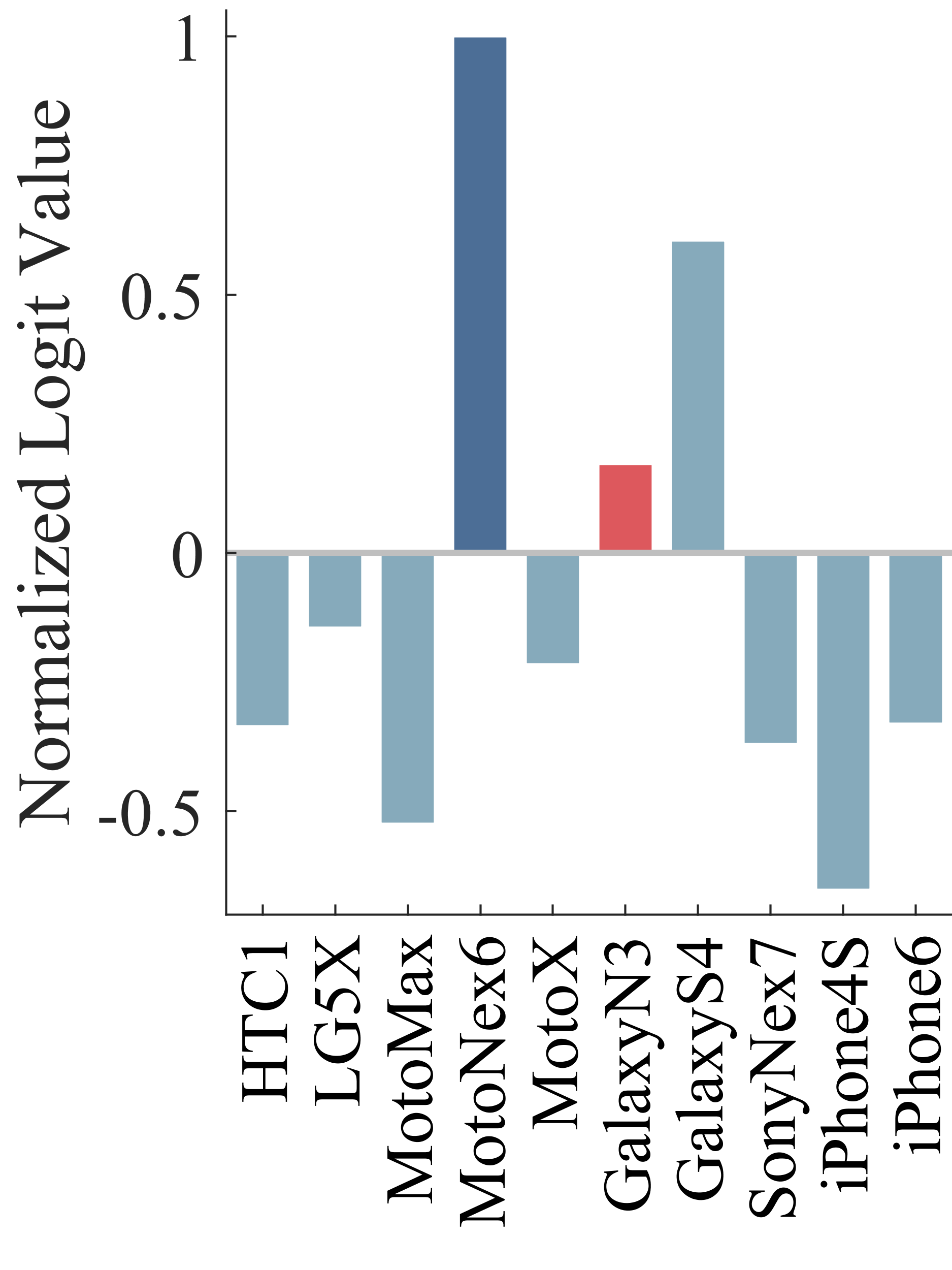}}
\end{minipage}
\end{minipage}
\end{center}
\vspace{-0.20cm}
\caption{An example of camera trace erasing in the classification task.
(a) Given an image, the state-of-the-art classifier \cite{ResNet-CMID} can predict the right image origin (red bar) with high confidence (measured by normalized logit value).
(b) A median filter effectively erases camera trace, which misleads the prediction (dark blue bar), yet at the cost of signal destruction.
(c) A real-world image denoising method \cite{TWSC}, though gently removing the visible noise, is not effective enough to erase camera trace.
(d) Our proposed method, SiamTE, effectively erases the camera trace without visible destruction of content signal. Zoom in for a better visual experience.}
\label{fig:intro}
\vspace{-0.27cm}
\end{figure}

Nevertheless, there is little systematic research on the performance of these forensic methods in an adversarial case. In this paper, we address camera trace erasing in order to reveal the weakness of trace-based forensic methods.
As an example shown in Fig.~\ref{fig:intro}, a median filter effectively degrades the classification accuracy of a forensic method \cite{ResNet-CMID}, yet at the cost of signal destruction. While a real-world image denoiser \cite{TWSC} gently removes the visible noise, the residual camera trace in the processed image is still sufficient for the classifier to make a right prediction.
Therefore, it is non-trivial to effectively erase camera trace while avoiding the destruction of content signal.

We propose Siamese Trace Erasing (SiamTE) to reconcile these two demands, which is implemented by a Convolutional Neural Network (CNN).
Specifically, we design a novel hybrid loss on the basis of Siamese architecture \cite{chopra2005learning} for network training.
This hybrid loss contains three terms: embedded similarity, truncated fidelity, and cross identity.
For the embedded similarity loss, we gather images captured by different types of cameras as a group of input, and measure the similarity between network outputs.
Theoretically, given a suitable metric for camera trace, the similarity of images captured by different types of cameras will increase along with the decrease of camera trace. It is thus possible to measure the degree of camera trace erasing by calculating this similarity.
Inspired by deep metric learning \cite{kulis2013metric}, we adopt the normalized Euclidean distance in a learned embedding space to be such a metric.

However, the embedded similarity loss alone is not sufficient to guide the network towards desired outputs, mainly due to the issue of over-manipulation.
We then introduce a truncated fidelity loss to restrict the degree of image manipulation, by minimizing the Manhattan distance between network inputs and outputs. More specifically, we truncate the distance values below a threshold to zero, which preserves essential manipulation for camera trace erasing while avoiding potential over-manipulation.
Besides the above two loss items, we further propose a cross identity loss for better disentanglement between camera trace and content signal.

To evaluate the anti-forensic performance of SiamTE, we conduct experiments on two datasets, in which images are captured by different types of cameras. We take three forensic tasks, \textit{i.e.}, classification, clustering, and verification into consideration. In the classification task, we adopt CNN-based methods for evaluation \cite{ResNet-CMID}. In the clustering task, we perform K-means clustering on CNN-extracted features \cite{Kmeans}. In the verification task, we adopt a classic forensic method using hand-crafted features \cite{CamFP}.
Compared with the existing methods, \textit{i.e.}, filtering, compression, denoising, deblocking, and gradient-based adversarial methods, SiamTE significantly boosts the anti-forensic performance in all three tasks, without visible destruction of content signal. It demonstrates a clear advantage of SiamTE for camera trace erasing.

Contributions of this paper are summarized as follows:
\vspace{-0.2cm}
\begin{itemize}[leftmargin=0.43cm] \itemsep -3.1pt

\item We address a new low-level vision problem, termed as camera trace erasing, to reveal the weakness of trace-based forensic methods.

\item We propose SiamTE as an advanced solution and design a novel hybrid loss based on Siamese architecture for network training.

\item SiamTE achieves a significant and consistent performance improvement over existing anti-forensic methods in terms of different datasets and tasks.

\end{itemize}

\section{Related Work}
\label{sec:related_work}

\begin{figure}[t]
\begin{center}
\includegraphics[width=0.99\linewidth]{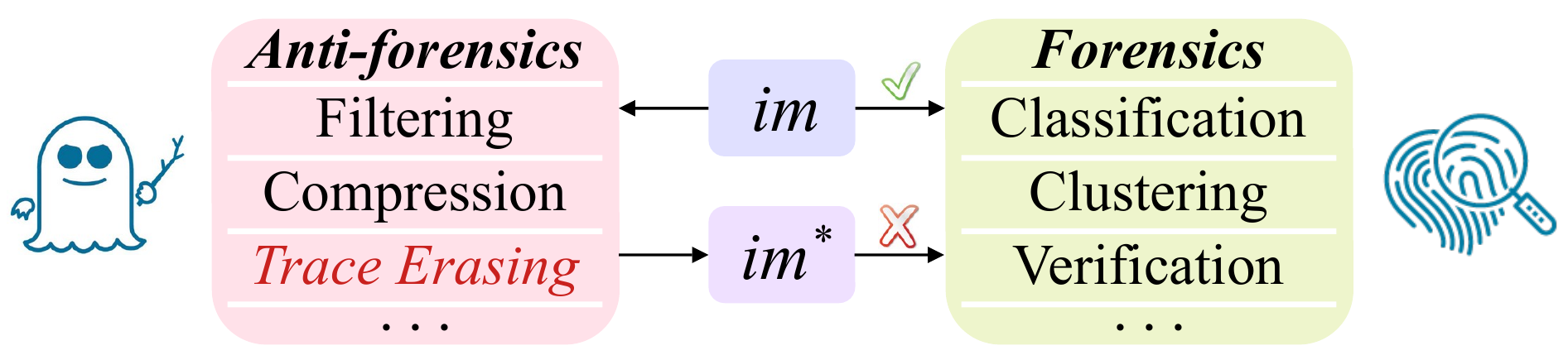}
\end{center}
\vspace{-0.38cm}
\caption{Image anti-forensics and image forensics. Forensics could fail when $im$ is processed to $im^*$ by anti-forensic methods.}
\label{fig:anti}
\vspace{-0.22cm}
\end{figure}

\textbf{Image anti-forensics}. In order to counter image forensics, researches have proposed a variety of anti-forensic methods to disguise the manipulation history of images (Fig.~\ref{fig:anti}). Among these methods, median filter and JPEG compression attract most research interests. Kirchner \etal adopt a median filter to hide traces of image resampling \cite{kirchner2008hiding}.
Though effective, the median filter itself will leave a distinctive pattern, termed as streaking artifact \cite{Streaking, Robust-median, Blind-median}. As a compensation, researches propose methods for streaking artifact removal, in order to disguise the manipulation history of median filter \cite{Chen-SPL2015, Kim-SPL2018}.
For JPEG compression, Fan \etal propose a variational approach to hide the traces (\textit{i.e.}, the blocking artifact) of compression \cite{Variational-jpeg}.
Furthermore, a dictionary-based method \cite{Dictionary-jpeg} and a GAN-based method \cite{Anti-forensics-jpeg-gan} are also proposed for JPEG anti-forensics.
In addition to deblocking after compression, researches propose anti-forensic methods in the JPEG compression process, by adding dithering to the transform coefficients \cite{Dither-jpeg, Anti-forensics-jpeg, Improved-jpeg}.

Camera trace erasing can be categorized to image anti-forensics. Compared with the existing settings, we make a step forward to address a more realistic trace. Unlike the streaking artifact caused by median filter and the blocking artifact caused by JPEG compression, it is difficult to conclude a fixed pattern for camera trace, since it varies from camera to camera.
Besides, we involve filtering, compression, and deblocking methods for comparison and conduct a comprehensive investigation to these anti-forensic methods in the problem of camera trace erasing.

\textbf{Adversarial technique}. The adversarial technique is designed to verify the security of defense system \cite{qiu2019review}. In this research direction, adversarial machine learning has attracted a lot of attention, due to the rapid development of learning-based methods \cite{kurakin2016adversarial}.
Researchers have discovered that an adversarial method can easily trigger malfunctions of a trained neural network, by adding gradient-based perturbation to network inputs \cite{goodfellow2018making, Analysis-adv, FGSM}.

Our proposed SiamTE can be viewed as an adversarial method, since it degrades the performance of trace-based forensic methods. But different from the common solutions, SiamTE works \textit{without} the support of gradient information, which makes it capable of handling more kinds of images, whose camera types are out of the known labels.

\textbf{Real-world image denoising}. Image denoising is a classic research topic in low-level computer vision. Recently, researchers have widened their focus from synthetic noise (\textit{e.g.,} Gaussian noise) \cite{Dabov-BM3D, Chen-TNRD, Zhang-DnCNN, N2V, DBF, MemNet} to the real-world ones \cite{CC, TWSC, SIDD, DBF-TPAMI, jourabloo2018face, N3Net}.
Since it is difficult to characterize a real-world noise produced by the complex and diverse in-camera processing pipelines, researcher have to define the noise-free image, in order to obtain the ground truth for evaluation. Two kinds of methods have been proposed for this definition, \textit{i.e.}, (a) multi-frame averaging \cite{CC, SIDD} and (b) paired-acquisition under low-ISO value \cite{DND}.

Camera trace can be viewed as one kind of real-world noise. However, the existing definitions of noise-free images are not suitable for camera trace erasing. For (a), researchers have found that a part of camera trace can still survive or even be enhanced after frame averaging \cite{holst1998ccd, PRNU}. For (b), an image cannot get rid of camera trace even if it is captured at a low-ISO setting. Thus, we directly define camera trace based on its characteristics. To the best of our knowledge, it is the first time that a specific definition is provided to a real-world noise. Besides, we adopt two representative real-world denoising methods for comparison and demonstrate the advantage of our proposed method.

\section{Siamese Trace Erasing}
\subsection{Problem formulation}
\label{sec:formulation}

We view a captured image ($im$ for short) as two parts. One is camera trace ($trs$ for short) and the other is content signal ($sig$ for short), which can be formulated as
\vspace{-0.1cm}
\begin{equation}
im = sig + trs.
\vspace{-0.1cm}
\end{equation}
The goal of a camera trace erasing method $F(\cdot)$ is to achieve $F(im) = sig$.
As detailed in Sec.~\ref{sec:related_work}, existing methods cannot provide a suitable definition to either the $sig$ part or the $trs$ part for a single image.

To address this issue, we propose to take multiple images into consideration at the same time. Firstly, we define that camera trace should be a distinguishable part in an image. In other words, a certain kind of camera trace should be different from the other one, when these two images are captured by different types of cameras.
From this definition, there shall exist a certain kind of similarity between images, which increases along with the decrease of the distinguishable part (\textit{i.e.}, camera trace) in each image.
Taking two images ($im_1$ and $im_2$) as an example, we denote $\phi(\cdot, \cdot)$ as the similarity between two images. With these notations, we have an inequation as
\vspace{-0.05cm}
\begin{equation}
\phi(F_1(im_1), F_1(im_2)) > \phi(F_2(im_1), F_2(im_2)),
\vspace{-0.05cm}
\label{eq:sim}
\end{equation}
when a trace erasing method $F_1(\cdot)$ is better than $F_2(\cdot)$.

Moreover, we define that camera trace should be the \textit{only} distinguishable part, regardless of the content signal. Ideally, when $sig_n=F(im_n), trs_n=im_n-F(im_n), n=1,2$, we have an equation as
\begin{equation}
\phi(sig_1+trs_2, sig_2+trs_1) = \phi(sig_1+trs_1, sig_2+trs_2).
\label{eq:cross}
\end{equation}
We use the above formulation to define camera trace and motivate our proposed trace erasing method.

\begin{figure}[t]
\begin{center}
\includegraphics[width=0.96\linewidth]{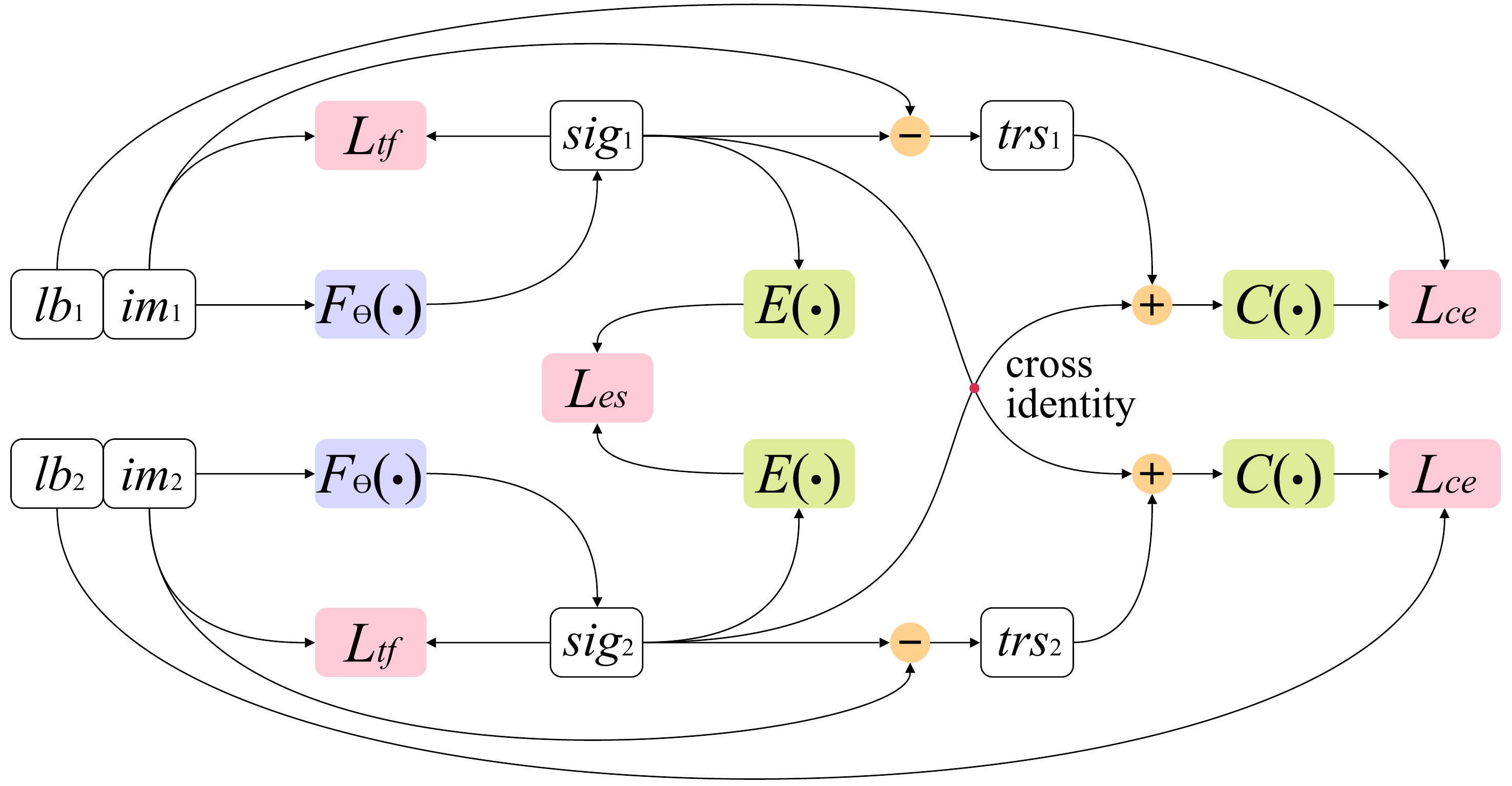}
\vspace{-0.35cm}
\end{center}
\caption{Flowchart of training strategy.
$im_n (n=1,2)$ denote images captured by the $n^{th}$ type of camera with labels $lb_n (n=1,2)$.
$sig_n$ and $trs_n$ denote the estimated content signal and camera trace, respectively.
$F_\Theta(\cdot)$ denotes a parametric method for camera trace erasing.
$E(\cdot)$ denotes a mapping for image embedding.
$C(\cdot)$ denotes a classier for image origin identification.
The hybrid loss contains embedded similarity $L_{es}$, truncated fidelity $L_{tf}$, and the cross identity part with cross-entropy loss $L_{ce}$.
\textbf{\textcolor{myBlue}{Blue}}: trainable model, \textbf{\textcolor{myGreen}{Green}}: fixed oracles, and \textbf{\textcolor{myRed}{Red}}: loss functions.
}
\label{fig:archi}
\vspace{-0.1cm}
\end{figure}

\subsection{Hybrid loss guided SiamTE}
\label{sec:method}

Let $F_\Theta(\cdot)$ denote a parametric method for camera trace erasing, where $\Theta$ is the trainable parameters.
In this paper, we adopt the CNN structure proposed in \cite{DBF-TPAMI} as an embodiment of $F_\Theta(\cdot)$. It is worth mentioning that, the focus of this paper is not the network design. $F_\Theta(\cdot)$ can be implemented by other network structures.

For the setup of network, we adopt the Siamese architecture proposed in \cite{chopra2005learning}. Taking two images as an example, we illustrate the Siamese architecture in Fig.~\ref{fig:archi}, where $F_\Theta(\cdot)$ is duplicated in two branches with shared parameters. Such a design can be easily generalized to multiple images, by adding more branches.
Following the name of architecture, we term our proposed method as Siamese Trace Erasing (\textit{SiamTE}). To train the network, we propose a hybrid loss as detailed below and shown in Fig.~\ref{fig:archi}.

\textbf{Embedded similarity loss} $L_{es}$. 
Motivated by inequation (\ref{eq:sim}), we propose the embedded similarity loss to guide the network training.
Inspired by deep metric learning \cite{kulis2013metric}, we adopt a learned metric to calculate the similarity between images, by embedding images to a trace-related space with $E(\cdot)$. In this embedding space, we calculate the normalized Euclidean distance between features to obtain the similarity.
The calculation process of embedded similarity loss is summarized in Algorithm~\ref{alg:1}.

\textbf{Truncated fidelity loss} $L_{tf}$. 
Generally, the intensity of camera trace is limited compared to content signal since it is a by-product in imaging process.
Motivated by this prior knowledge, we propose truncated fidelity loss to restrict the manipulation of a camera trace method $F_\Theta(\cdot)$. Specifically, we calculate the Manhattan distance between $im$ and $F_\Theta(im)$ to measure the degree of manipulation.
To preserve essential manipulation while avoiding potential over-manipulation, we truncate the distance values below a threshold $T$ to zero, which can be formulated as
\vspace{-0.05cm}
\begin{equation}
L_{tf} =
\begin{cases}
|im - F_\Theta(im)|, &|im - F_\Theta(im)| > T \\
0, & |im - F_\Theta(im)| \leq T
\end{cases}.
\vspace{-0.05cm}
\end{equation}

\textbf{Cross identity loss} $L_{ci}$.
Motivated by equation (\ref{eq:cross}), we propose cross identity loss for better disentanglement between camera trace and content signal.
Suppose we have $G$ types of cameras, images captured from different  types of cameras are grouped as network input $im_n, n=1,2,...,G$.
Let $trs_m$ denotes a camera trace produced by the $m^{th}$ camera. The cross identity loss aims to maximize the probability of synthetic images $sig_n + trs_m, n=1,2,...,G (n \neq m)$ to be identified as the ones captured by the $m^{th}$ camera device.
In our implementation, we combine an estimated trace $im_m-F_\Theta(im_m)$ with signals from the other devices $F_\Theta(im_n), n=1,2,...,G (n \neq m)$, to obtain these synthetic images.
Then, we forward them into a trained classifier $C(\cdot)$ to obtain feedback.
The calculation process of cross identity loss is summarized in Algorithm~\ref{alg:2}.

\begin{algorithm}[t]
\caption{Calculation of embedded similarity loss}
\label{alg1}
\begin{algorithmic}[1]
\Require\\
$im[\cdot]$: $G$ images captured by different cameras;\\
$F_\Theta(\cdot)$: a method for trace erasing with parameters $\Theta$;\\
$E(\cdot)$: a learned mapping for image embedding; \\
$N(\cdot)$: an operator for $L_2$ normalization; \\
$S(\cdot,k)$: an operator for cyclic shift with $k$ step size; \\
$D(\cdot,\cdot)$: an operator to calculate Euclidean distance; \\
$M$: a margin for Euclidean distance.

\Ensure embedded similarity loss of $im$, $L_{es}$
\For {each $g \in [1, G]$}
\State Trace erasing: $sig[g] = F_{\Theta}(im[g])$;
\State Embedding: $feat[g] = E(sig[g])$;
\State Normalization: $feat[g] = N(feat[g])$;
\EndFor \\
Initialize $L_{es}$ to zero;
\For{$k=1$; $k<G$; $k=k+1$}
\State $dist = \ $max$(0, D(feat, S(feat, k)) - M)$;
\State $L_{es} = L_{es}\ + \ $mean$(dist)$;
\EndFor \\
Average: $L_{es} = L_{es} / (G-1)$; \\
\Return $L_{es}$
\end{algorithmic}
\label{alg:1}
\end{algorithm}

\begin{figure}[t]
\begin{center}
\includegraphics[width=0.8\linewidth]{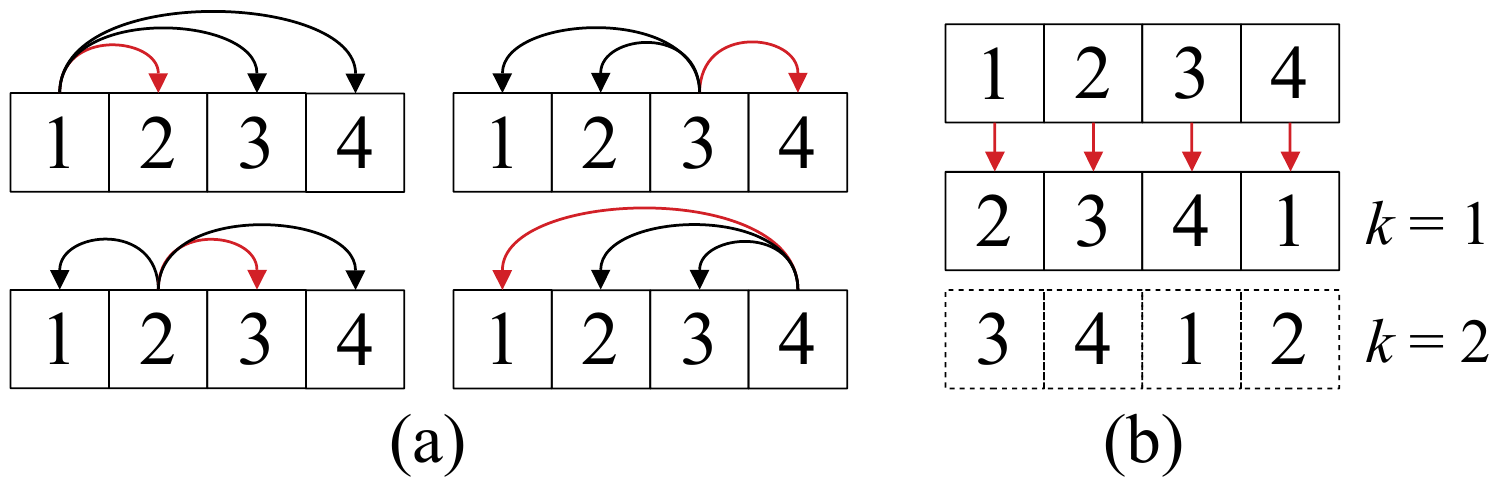}
\end{center}
\vspace{-0.35cm}
\caption{Calculating one-to-one operation in a parallel computing friendly way.
Given $G$ images in a group ($G=4$ in this example), the number of sequential executions can be reduced from (a) $\mathrm{P}_G^2$ (or $\mathrm{C}_G^2$ when an operation is commutative) to (b) $G-1$. With the cyclic shift $S(\cdot,k)$, the operations colored in \textcolor{red}{red} arrows can be calculated in parallel. For each execution, the step size of cyclic shift $k$ traverses from $1$ to $G-1$.}
\label{fig:cycle_shift}
\vspace{-0.1cm}
\end{figure}

\subsection{Implementation details}

In Algorithms~\ref{alg:1} and ~\ref{alg:2}, we adopt a cyclic shift operator $S(\cdot,k)$ with a step size of $k$, in order to calculate the one-to-one operation between multiple images in a parallel computing friendly way. We illustrate the operation of cyclic shift in Fig.~\ref{fig:cycle_shift}.
For the image origin classifier $C(\cdot)$, we train a ResNet \cite{ResNet} on KCMI+ (a dataset detailed in Sec.~\ref{sec:dataset}). The weights of convolutions in ResNet are initialized by an ImageNet pretrained model \cite{PyTorch}. After training, we utilize the stacked convolutions in this network as the embedding function $E(\cdot)$ for $L_{es}$.
Lastly, we linearly combine the above three loss functions to form the hybrid loss as $\lambda_1 L_{es} + \lambda_2 L_{tf} + \lambda_3 L_{ci}$, where $\lambda_{n} (n=1,2,3)$ denote the weighting factor.

\begin{algorithm}[t]
\caption{Calculation of cross identity loss}
\label{alg2}
\begin{algorithmic}[1]
\Require\\
$im[\cdot]$: $G$ images captured by different cameras; \\
$lb[\cdot]$: the corresponding origin labels of $im[\cdot]$; \\
$F_\Theta(\cdot)$: a method for trace erasing with parameters $\Theta$; \\
$C(\cdot)$: a method for image origin classification; \\
$S(\cdot,k)$: an operator for cyclic shift with $k$ step size; \\
$L_{ce}(\cdot, \cdot)$: an operator to calculate cross-entropy loss.

\Ensure cross identity loss of $im$, $L_{ci}$
\For {each $g \in [1, G]$}
\State Trace extraction: $trs[g] = im[g] - F_{\Theta}(im[g])$;
\EndFor \\
Initialize $L_{ci}$ to zero;
\For{$k=1$; $k<G$; $k=k+1$}
\State Cross identity: $pred = C(F_{\Theta}(im) + S(trs, k))$;
\State $L_{ci} = L_{ci}\ + \ $mean$(L_{ce}(pred, S(lb, k)))$;
\EndFor \\
Average: $L_{ci} = L_{ci} / (G-1)$; \\
\Return $L_{ci}$
\end{algorithmic}
\label{alg:2}
\end{algorithm}

\section{Experiments and Results}

\subsection{Datasets and settings}
\label{sec:dataset}
\textbf{KCMI}. 
Kaggle Camera Model Identification (KCMI) is a dataset proposed by IEEE’s Signal Processing Society \cite{KCMI}. In KCMI, $2,750$ images are captured with $10$ types of cameras. We separate $550$ images from it to build a test set KCMI-550, with $55$ images for each camera.
For training and validation, we first retrieve and download additional $2,800$ images from Flickr, which are captured with the same $10$ types of cameras. Then, we combine them with the rest $2,200$ images in KCMI to build KCMI+, with $500$ images for each camera.

\textbf{VISION}.
It is a large-scale dataset for source identification \cite{VISION}, in which images are captured by $30$ types of cameras. Among these cameras, $28$ are different from those in KCMI.
We adopt $1,500$ images from VISION to build a test set VISION-1500, with $50$ images for each camera.

\begin{figure*}
\fontsize{8}{9.6}\selectfont
\begin{center}
\begin{minipage}{0.96\linewidth}
\begin{minipage}{0.245\linewidth}
  \centerline{\includegraphics[width=1\linewidth]{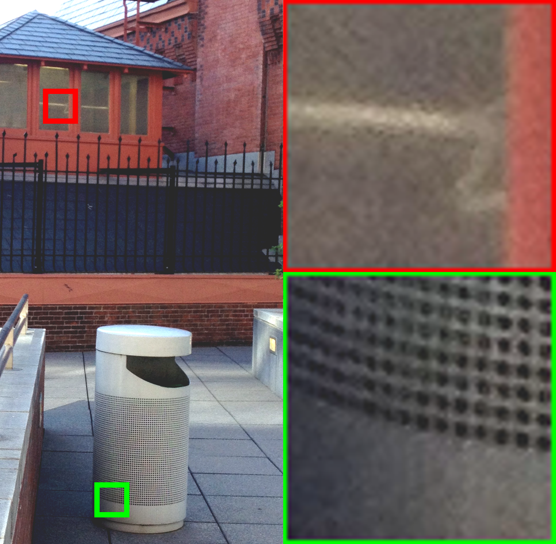}}
  \centerline{(a) ORI}
\end{minipage}
\hfill
\begin{minipage}{0.245\linewidth}
  \centerline{\includegraphics[width=1\linewidth]{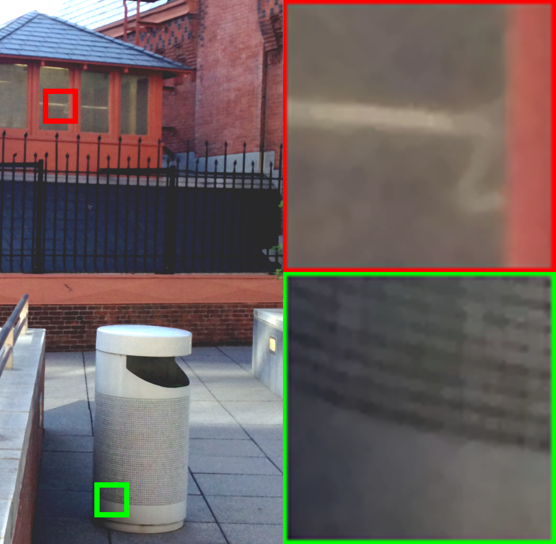}}
  \centerline{(b) MF5}
\end{minipage}
\hfill
\begin{minipage}{0.245\linewidth}
  \centerline{\includegraphics[width=1\linewidth]{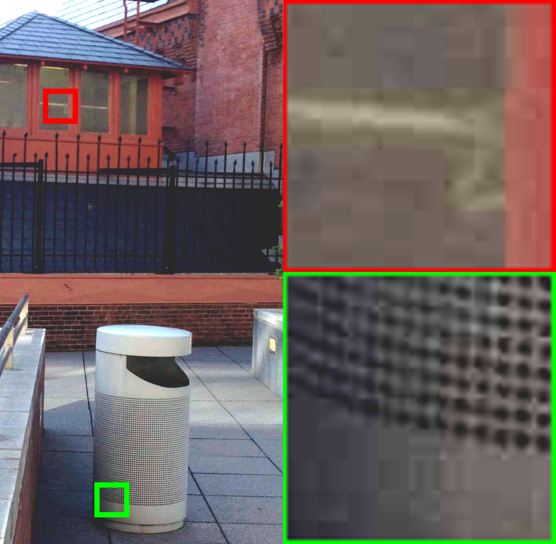}}
  \centerline{(c) CP30}
\end{minipage}
\hfill
\begin{minipage}{0.245\linewidth}
  \centerline{\includegraphics[width=1\linewidth]{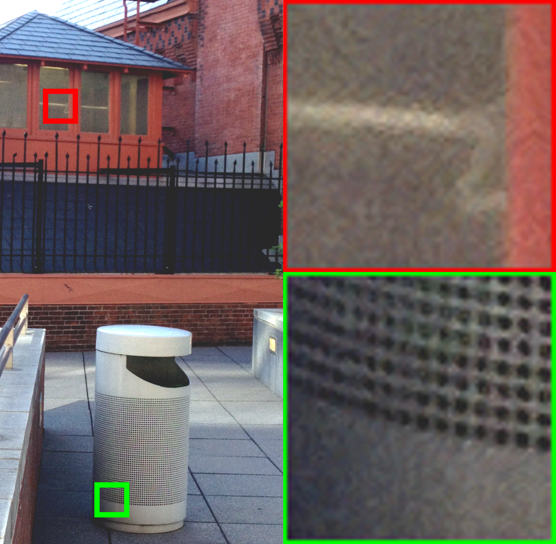}}
  \centerline{(d) AD2}
\end{minipage}
\vfill
\begin{minipage}{0.245\linewidth}
  \centerline{\includegraphics[width=1\linewidth]{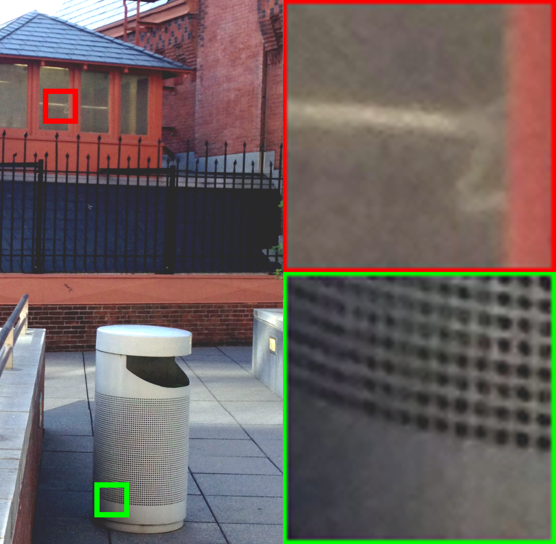}}
  \centerline{(e) DB}
\end{minipage}
\hfill
\begin{minipage}{0.245\linewidth}
  \centerline{\includegraphics[width=1\linewidth]{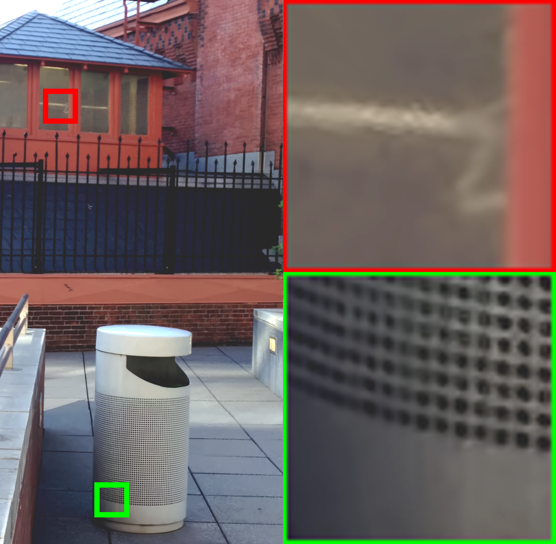}}
  \centerline{(f) DN-I}
\end{minipage}
\hfill
\begin{minipage}{0.245\linewidth}
  \centerline{\includegraphics[width=1\linewidth]{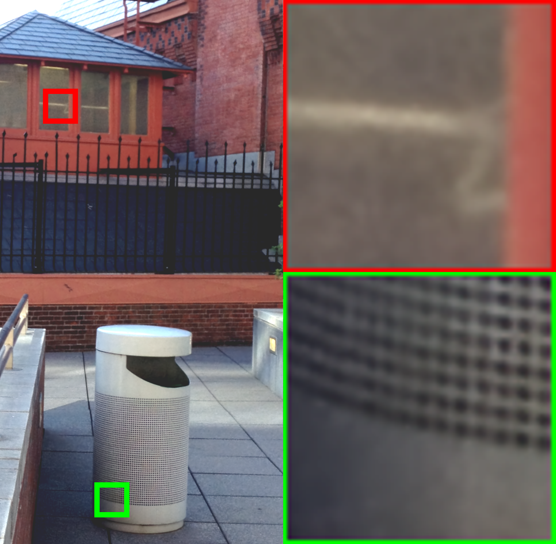}}
  \centerline{(g) DN-E}
\end{minipage}
\hfill
\begin{minipage}{0.245\linewidth}
  \centerline{\includegraphics[width=1\linewidth]{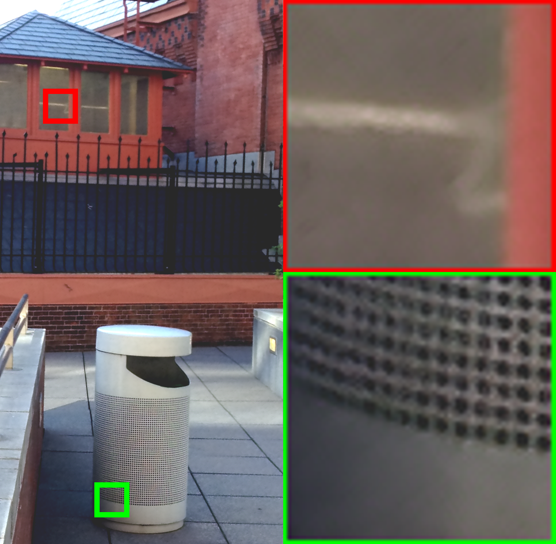}}
  \centerline{(h) Ours}
\end{minipage}
\end{minipage}
\end{center}
\vspace{-0.18cm}
\caption{Visual comparison on an image from KMCI-550.}
\vspace{-0.25cm}
\label{fig:vcomp1}
\end{figure*}

\textbf{Settings of training}.
We adopt KCMI+ to train SiamTE. Images from KCMI+ are randomly copped into patches with a size of $336\times336$. We randomly gather $4$ patches as a group (\textit{i.e.}, $G=4$), in which patches are cropped from images captured by different  types of cameras. $64$ groups are randomly gathered into a mini-batch for the stochastic gradient descent.
We adopt Adam \cite{Adam} for training with the momentum factor set as $0.9$. 
In our implementations, we set $\lambda_1$:$\lambda_2$:$\lambda_3$ as $3$:$1000$:$1$ or $3$:$500$:$1$ for the hybrid loss.
For the hyper-parameters, we set the margin $M=0.5$ and the threshold $T=3$.

\subsection{Image forensic tasks and metrics}
\label{sec:forensic_tasks_metrics}

\textbf{Classification}.
In this task, given an input image, the forensic methods predict a camera type as output. We adopt two kinds of CNN-based classifiers proposed in \cite{ResNet-CMID} for evaluation, named as ResNet50 and DenseNet201 according to respective network structures.
Since the input size of classification networks ($224\times224$) is much smaller than that of a full image, we randomly crop $4$ patches in an image as its representative, and choose the majority prediction among $4$ patches as the final output.
Since the classification accuracy varies with different cropped patches, we repeat each experiment $10$ times and report the averaged results. KCMI-550 is adopted for evaluation.

\begin{table}
\begin{center}
\caption{Quantitative comparison in the classification task.}
\label{tab:classification}
\begin{threeparttable}
\fontsize{8}{9.6}\selectfont
\begin{tabular}{L{1.15cm}C{1.70cm}C{1.70cm}C{0.75cm}C{0.75cm}}
\toprule[1.2pt]

\multirow{2}{*}{Method} & \multicolumn{2}{c}{Accuracy} & \multirow{2}{*}{NIQE} & \multirow{2.5}{*}{\tabincell{c}{$L_1$ dist.\\to ORI}} \\
\cmidrule(lr){2-3}
& ResNet50 & DenseNet201\\

\midrule
ORI	\tnote{1}			& $99.80\pm0.18$ & $99.87\pm0.12$ & $3.083$ & - \\
\midrule
MF3 \cite{Chen-SPL2015}			& $66.62\pm2.08$ & $75.95\pm1.43$ & $3.939$ & $2.055$ \\ 
MF5 \cite{Chen-SPL2015}			& $21.33\pm1.09$ & $44.33\pm1.64$ & $4.799$ & $3.849$ \\
GF3 \cite{russ2016forensic}				& $90.18\pm0.66$ & $93.07\pm1.02$ & $4.256$ & $2.186$ \\
GF5 \cite{russ2016forensic}				& $73.46\pm1.56$ & $80.18\pm1.29$ & $4.593$ & $3.115$ \\
\midrule
CP30 \cite{jpeg}			& $56.31\pm1.64$ & $58.35\pm1.06$ & $4.283$ & $3.632$ \\
CP40 \cite{jpeg}			& $77.02\pm1.51$ & $75.27\pm1.22$ & $3.838$ & $3.202$ \\
CP50 \cite{jpeg}			& $91.29\pm0.73$ & $86.56\pm1.37$ & $3.510$ & $2.917$ \\
\midrule
AD1 \cite{FGSM}			& $45.49\pm1.43$ & $55.53\pm1.34$ & $3.265$ & $0.988$ \\
AD2 \cite{FGSM}			& $21.13\pm1.25$ & $32.73\pm1.53$ & $3.934$ & $1.973$ \\
\midrule
DB \cite{Zhang-DnCNN} 				& $90.04\pm1.03$ & $92.82\pm1.05$ & $3.044$ & $1.301$ \\
DN-I \cite{TWSC}			& $59.49\pm1.79$ & $64.87\pm1.06$ & $3.961$ & $2.017$ \\
DN-E \cite{DBF-TPAMI}			& $44.42\pm1.66$ & $56.82\pm1.32$ & $4.008$ & $2.710$ \\
\midrule
Ours			& $\mathbf{20.42}\pm1.19$ & $\mathbf{28.11}\pm1.76$ & $3.676$ & $2.004$ \\
\bottomrule[1.2pt]
\end{tabular}
\begin{tablenotes}
\footnotesize
	\item[1] Images from KCMI-550 (ORI) are adopted for evaluation.
\end{tablenotes}
\end{threeparttable}
\end{center}
\vspace{-0.7cm}
\end{table}

\begin{table*}
\begin{center}
\caption{Quantitative comparison in the clustering task. Images from VISION-1500 (ORI) are adopted for evaluation.}
\label{tab:clustering}
\begin{threeparttable}
\fontsize{8}{9.6}\selectfont
\begin{tabular}{L{1.15cm}C{1.70cm}C{1.70cm}C{1.70cm}C{1.70cm}C{1.70cm}C{1.70cm}C{0.8cm}C{0.8cm}}
\toprule[1.2pt]
\multirow{3}{*}{Method \tnote{1}} & \multicolumn{6}{c}{Accuracy} & \multirow{3}{*}{NIQE} & \multirow{3.5}{*}{\tabincell{c}{$L_1$ dist.\\to ORI}} \\
\cmidrule(lr){2-7}
& ResNet50 ($K=30$) \tnote{2} & ResNet50 ($K=60$) & ResNet50 ($K=90$) & DenseNet201 ($K=30$) \tnote{2} & DenseNet201 ($K=60$) & DenseNet201 ($K=90$) \\

\midrule
ORI			& $56.79\pm2.08$ & $70.27\pm1.01$ & $75.44\pm1.30$ & $59.06\pm1.79$ & $73.09\pm1.27$ & $78.77\pm1.53$ & $3.585$ & - \\
\midrule
MF3 \cite{Chen-SPL2015}		& $43.03\pm1.36$ & $54.42\pm1.15$ & $58.60\pm1.08$ & $45.23\pm1.14$ & $55.73\pm1.13$ & $61.79\pm1.32$ & $4.043$ & $1.959$ \\
MF5 \cite{Chen-SPL2015}		& $31.45\pm1.06$ & $38.53\pm0.63$ & $42.79\pm1.15$ & $34.97\pm1.02$ & $42.79\pm0.98$ & $48.14\pm0.76$ & $5.227$ & $3.525$ \\
GF3 \cite{russ2016forensic} 		& $49.95\pm1.74$ & $61.68\pm1.55$ & $64.99\pm1.25$ & $51.17\pm0.98$ & $61.02\pm1.74$ & $66.19\pm0.86$ & $4.413$ & $2.029$ \\
GF5 \cite{russ2016forensic}		& $44.69\pm1.27$ & $51.82\pm0.93$ & $55.87\pm0.56$ & $41.47\pm1.49$ & $52.43\pm1.56$ & $57.40\pm1.02$ & $4.795$ & $2.847$ \\
\midrule
CP30 \cite{jpeg} 		& $26.55\pm1.06$ & $35.32\pm1.55$ & $39.41\pm0.74$ & $25.23\pm1.08$ & $33.27\pm0.80$ & $37.68\pm1.12$ & $4.811$ & $3.564$ \\
CP40 \cite{jpeg} 		& $33.57\pm1.62$ & $43.45\pm1.61$ & $47.82\pm1.60$ & $31.44\pm1.13$ & $40.28\pm0.89$ & $44.12\pm1.10$ & $4.253$ & $3.175$ \\
CP50 \cite{jpeg} 		& $40.91\pm1.30$ & $49.62\pm1.76$ & $53.70\pm0.95$ & $35.95\pm1.24$ & $45.78\pm0.84$ & $50.87\pm1.19$ & $3.965$ & $2.918$ \\
\midrule
DB \cite{Zhang-DnCNN} 			& $52.87\pm1.61$ & $64.13\pm1.68$ & $70.22\pm1.56$ & $53.62\pm1.92$ & $67.44\pm0.90$ & $72.44\pm1.64$ & $3.299$ & $1.327$ \\
DN-I \cite{TWSC}		& $35.83\pm1.11$ & $46.44\pm0.87$ & $51.23\pm0.73$ & $37.06\pm1.09$ & $46.31\pm1.17$ & $51.84\pm0.81$ & $4.275$ & $2.214$  \\
DN-E \cite{DBF-TPAMI}		& $28.47\pm0.89$ & $37.83\pm1.40$ & $42.64\pm0.86$ & $28.45\pm1.01$ & $37.76\pm0.71$ & $43.71\pm0.82$ & $4.128$ & $2.613$ \\
\midrule
Ours		& $\mathbf{23.44}\pm0.82$ & $\mathbf{33.37}\pm1.06$ & $\mathbf{37.30}\pm0.83$ & $\mathbf{22.94}\pm0.97$ & $\mathbf{31.99}\pm1.05$ & $\mathbf{37.24}\pm0.98$ & $4.082$ & $2.097$ \\
\bottomrule[1.2pt]
\end{tabular}
\begin{tablenotes}
\footnotesize
	\item[1] AD is not involved for comparison, since it cannot generalize to images with unseen camera types, as detailed in Sec.~\ref{sec:anti-forensics}.
	\item[2] Features extracted by the stacked convolutional layers are adopted for K-means clustering \cite{Kmeans}.
\end{tablenotes}
\end{threeparttable}
\end{center}
\vspace{-0.40cm}
\end{table*}

\begin{figure*}
\fontsize{8}{9.6}\selectfont
\begin{center}
\begin{minipage}{0.96\linewidth}
\begin{minipage}{0.245\linewidth}
  \centerline{\includegraphics[width=1\linewidth]{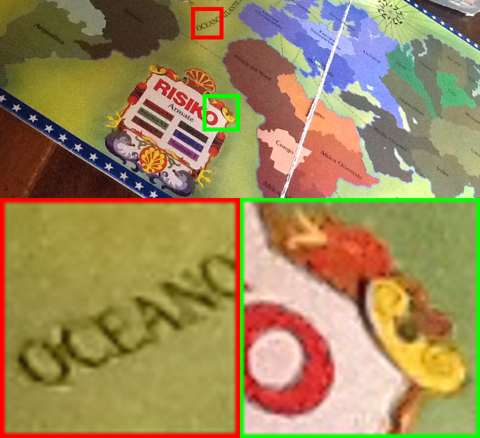}}
  \centerline{(a) ORI}
\end{minipage}
\hfill
\begin{minipage}{0.245\linewidth}
  \centerline{\includegraphics[width=1\linewidth]{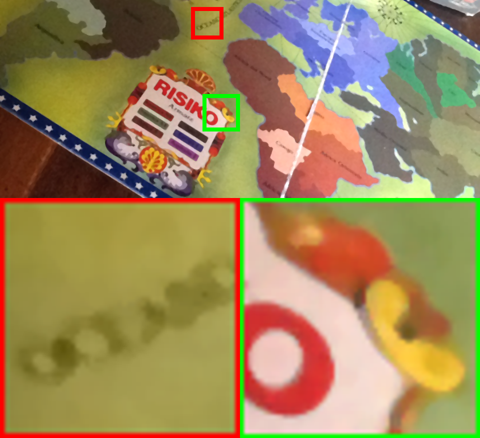}}
  \centerline{(b) MF5}
\end{minipage}
\hfill
\begin{minipage}{0.245\linewidth}
  \centerline{\includegraphics[width=1\linewidth]{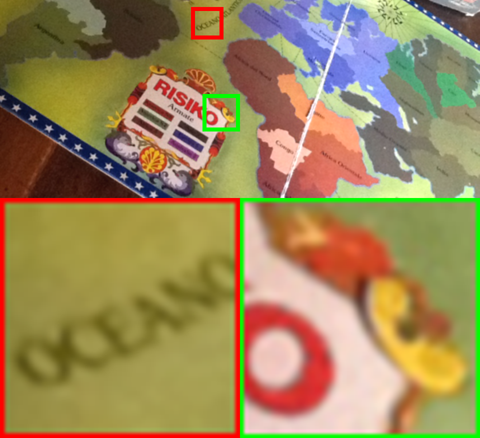}}
  \centerline{(c) GF5}
\end{minipage}
\hfill
\begin{minipage}{0.245\linewidth}
  \centerline{\includegraphics[width=1\linewidth]{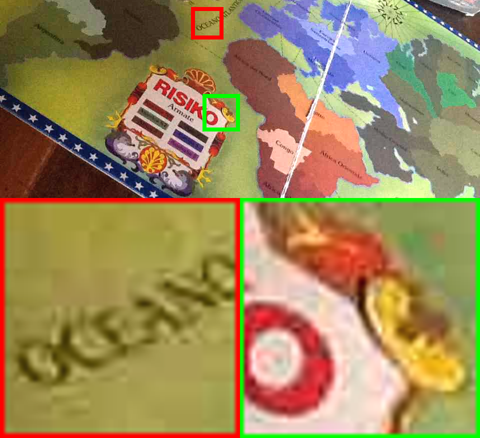}}
  \centerline{(d) CP30}
\end{minipage}
\vfill
\begin{minipage}{0.245\linewidth}
  \centerline{\includegraphics[width=1\linewidth]{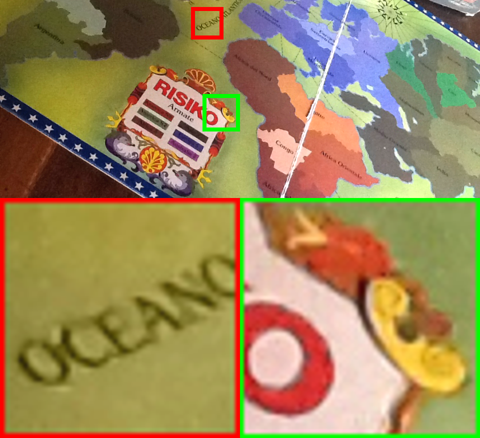}}
  \centerline{(e) DB}
\end{minipage}
\hfill
\begin{minipage}{0.245\linewidth}
  \centerline{\includegraphics[width=1\linewidth]{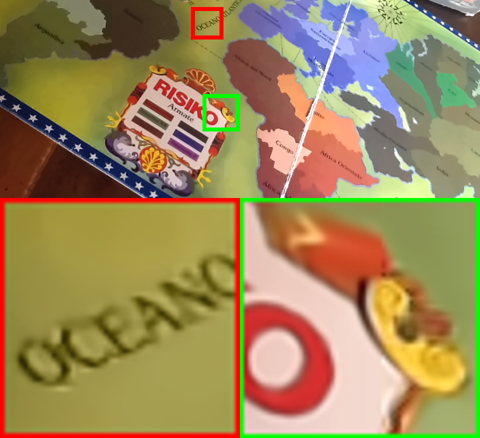}}
  \centerline{(f) DN-I}
\end{minipage}
\hfill
\begin{minipage}{0.245\linewidth}
  \centerline{\includegraphics[width=1\linewidth]{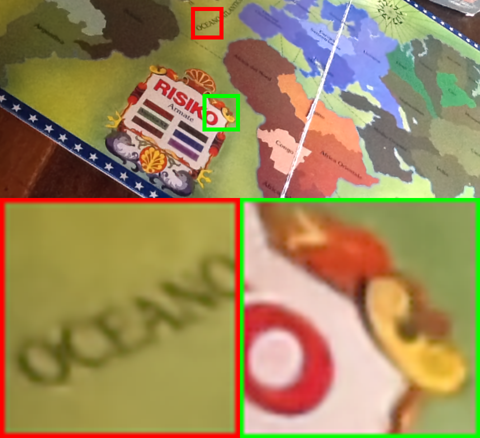}}
  \centerline{(g) DN-E}
\end{minipage}
\hfill
\begin{minipage}{0.245\linewidth}
  \centerline{\includegraphics[width=1\linewidth]{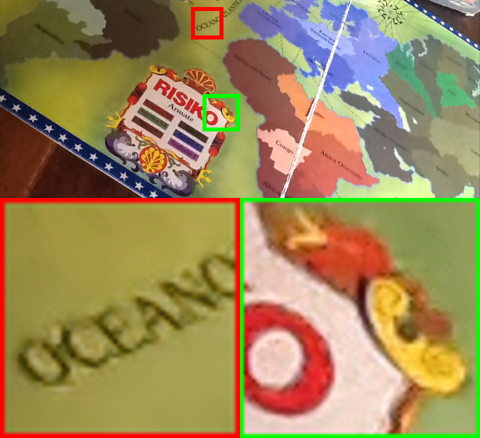}}
  \centerline{(h) Ours}
\end{minipage}
\end{minipage}
\end{center}
\vspace{-0.15cm}
\caption{Visual comparison on an image from VISION-1500.}
\label{fig:vcomp2}
\vspace{-0.25cm}
\end{figure*}

\textbf{Clustering}.
Besides the classification task conducted on KCMI-550, we perform clustering on VISION-1500 to evaluate the generalization ability of our proposed method.
Specifically, for a SiamTE trained on KCMI+, most of camera types on VISION-1500 are \textit{unseen}\footnote{The \textit{unseen} camera type is the one not contained in the training set (\textit{i.e.}, \textit{unknown} to methods), but we still know its origin.} during training. Thus, clustering is a more challenging task.
We adopt the stacked convolutional layers in ResNet50 and DenseNet201 \cite{ResNet-CMID} for feature extraction, and perform K-means clustering on extracted features \cite{Kmeans}. Since VISION-1500 contains $30$ types of cameras, we set the number of clustering center $K=30, 60,$ and $90$ for evaluation, respectively.
In order to provide quantitative results for clustering performance, we define an accuracy for clustering. Specifically, in a group of clustered images, we assign the majority camera type to them as predictions and compare these predictions with known image origins.
Therefore, such a clustering accuracy will decrease when images with different camera types are wrongly clustered together.
We randomly crop $128$ patches in an image as its representative and repeat each experiment $10$ times to obtain the averaged results.

\begin{table*}
\begin{center}
\caption{Quantitative comparison in the verification task. Images from KCMI-550 (ORI) are adopted for evaluation.}
\label{tab:identification}
\begin{threeparttable}
\fontsize{8}{9.6}\selectfont
\begin{tabular}{L{1.15cm}C{0.79cm}C{0.79cm}C{0.79cm}C{0.79cm}C{0.79cm}C{0.79cm}C{0.79cm}C{0.79cm}C{0.79cm}C{0.79cm}C{0.79cm}C{0.79cm}C{0.79cm}}
\toprule[1.2pt]
\multirow{2}{*}{Method} & \multicolumn{11}{c}{Peak-to-Correlation Energy} & \multirow{2}{*}{NIQE} & \multirow{2.5}{*}{\tabincell{c}{$L_1$ dist.\\to ORI}} \\
\cmidrule(lr){2-12}
& 1 & 2 & 3 & 4 & 5 & 6 & 7 & 8 & 9 & 10 & Avg. \tnote{1} \\
\midrule
ORI  	& $647.7$	&	$390.7$	&	$1604.0$		&	$300.2$	&	$2068.2$	&	$1388.6$	&	$1170.4$	&	$11.9$	&	$2896.0$		&	$2592.8$	&	$1307.1$ & $3.772$ & - \\
\midrule
MF3 \cite{Chen-SPL2015} 	& $236.7$ 	& $133.9$ 	& $630.8$ 		& $119.9$ 	& $623.6$ 	& $612.2$ 	& $400.6$ 	& $5.9$ 	& $686.4$ 		& $921.8$ 	& $437.1$ & $4.699$ & $2.325$ \\
GF3 \cite{russ2016forensic}		& $411.9$	&	$275.6$	&	$1143.9$		& $271.1$	&	$1290.1$	&	$1066.1$	&	$897.5$	&	$9.0$	&	$1850.8$		&	$1717.5$	& $893.3$ & $4.589$ & $2.521$ \\
CP50 \cite{jpeg}	& $65.7$		&	$58.4$		&	$250.3$		&	$61.1$		&	$274.4$	&	$144.0$	&	$148.7$	&	$2.5$	&	$343.9$		&	$493.5$	&	$184.3$ & $4.070$ & $3.104$ \\
AD2 \cite{FGSM} 	& $321.4$	&	$196.5$	& $1061.8$		&	$152.2$	&	$1118.9$	&	$637.7$	&	$643.8$	&	$7.1$	&	$1802.2$		&	$1568.0$	&	$751.0$ & $4.855$ & $1.994$ \\
DB \cite{Zhang-DnCNN} 		& $399.2$	& $267.4$ 	& $1197.9$ 	& $229.8$ 	& $1407.9$	& $877.0$	& $862.0$	& $8.2$	& $2388.6$		& $1740.8$	& $937.9$ & $3.832$ & $1.312$ \\
DN-I \cite{TWSC}	& $226.5$	&	$147.0$	&	$553.4$		&	$58.7$		&	$339.4$	&	$418.0$	&	$311.3$	&	$5.5$	&	$183.9$		&	$471.2$	&	$271.5$ & $4.600$ & $2.097$ \\
DN-E \cite{DBF-TPAMI}	& $263.0$	&	$198.8$	&	$854.7$		&	$209.6$	&	$989.5$	&	$584.3$	&	$682.7$	&	$6.3$	&	$1613.2$		&	$1219.6$	&	$662.2$ &$4.801$& $3.007$ \\
\midrule
Ours			& $53.7$ & $66.9$ & $107.1$ & $74.1$ & $271.9$ & $74.9$ & $233.4$ & $2.1$ & $410.6$ & $266.5$ & $\mathbf{156.1}$  & $4.652$ & $2.491$ \\
\bottomrule[1.2pt]
\end{tabular}
\begin{tablenotes}
\footnotesize
	\item[1] $10$ camera types represented by serial numbers are described in the supplementary document. Images are centrally cropped to simplify the calculation.
\end{tablenotes}
\end{threeparttable}
\end{center}
\vspace{-0.5cm}
\end{table*}

\textbf{Verification}.
Furthermore, we conduct evaluation in the verification task. Given two input images, the forensic method in this task predicts if they are captured by the same type of camera. We adopt KCMI+ to build fingerprints for each type of camera, by averaging the extracted noise residuals from images captured by the same type of camera. We adopt a hand-crafted method to extract the noise residual \cite{CamFP}.
Then, we adopt Peak-to-Correlation Energy (PCE) \cite{PCE} to measure the correlation between an image from KCMI-550 and its corresponding camera fingerprint. The higher PCE means a forensic method can identify the image origin with higher confidence \cite{kay1993fundamentals}.

\textbf{Metrics for image assessment}.
In addition to the specific metrics for each task, we provide an auxiliary metric to measure the manipulation degree of a trace erasing method, by calculating the Manhattan ($L_1$) distance between input and output of the method.
Generally, a large $L_1$ distance indicates that the input image has been heavily manipulated, which usually causes a destruction to content signal. While a small $L_1$ distance may indicate that a method fails to perform a valid manipulation on the input.
Moreover, we conduct objective quality assessments using the non-reference metric NIQE \cite{NIQE} (a smaller value denotes a higher quality).

\vspace{-0.05cm}
\subsection{Evaluation on anti-forensics performance}
\vspace{-0.05cm}
\label{sec:anti-forensics}

We conduct a comprehensive investigation on various existing anti-forensic methods for camera trace erasing, including median filter (MF), Gaussian filter (GF), JPEG compression (CP), gradient-based adversarial method (AD) \cite{FGSM}, blind image deblocking (DB) \cite{Zhang-DnCNN}, internal similarity based real-world denoising (DN-I) \cite{TWSC}, and learning-based real-world denoising (DN-E) \cite{DBF-TPAMI}.
The number after MF and GF (\textit{i.e.}, $3$ and $5$) denotes the kernel size of filter, the number after CP (\textit{i.e.}, $30$, $40$, and $50$) denotes the quality factor, and the number after AD (\textit{i.e.}, $1$ and $2$) denotes the scale factor of adversarial dithering.

As described in Sec.~\ref{sec:forensic_tasks_metrics}, we adopt the $L_1$ distance to measure the degree of image manipulation. Taking the results of CP in Table~\ref{tab:classification} as an example, a better anti-forensic performance is achieved with a larger degree of manipulation (\textit{e.g.,} CP30 vs.\ CP50). However, such a performance improvement comes at the cost of more severe signal destruction (reflected by a larger NIQE value).
Thus, to measure the \textit{efficiency} of a camera trace erasing method, we need to consider the degree of manipulation at the same time. An efficient method should achieve good anti-forensic performance with as little manipulation as possible.

According to the quantitative results listed in Tables~\ref{tab:classification}, \ref{tab:clustering}, \ref{tab:identification} and visual results shown in Figs.~\ref{fig:vcomp1} and \ref{fig:vcomp2}, we analyze the performance of above mentioned anti-forensic methods.
Median filter and JPEG compression are effective to erase camera trace, yet the processed images suffer from blurring and blocking artifacts. Gaussian filter is less effective since it cannot significantly degrade the performance of forensic methods even at a large degree of manipulation. The deblocking method \cite{Zhang-DnCNN} cannot provide a valid manipulation to erase camera trace, since it can only remove the part of blocking artifact in camera trace.
The overall performance of the two real-world image denoising methods \cite{TWSC, DBF-TPAMI} is relatively better than other baseline methods, yet still has a notable gap to ours.

\begin{table}
\begin{center}
\caption{Ablation study of the hybrid loss.}
\label{tab:ablation}
\begin{threeparttable}
\fontsize{8}{9.6}\selectfont
\begin{tabular}{L{0.25cm}L{0.25cm}L{0.25cm}C{1.55cm}C{1.55cm}C{0.75cm}C{0.75cm}}
\toprule[1.2pt]
\multicolumn{3}{c}{Hybrid Loss} & \multicolumn{2}{c}{Accuracy \tnote{1}} & \multirow{2}{*}{NIQE} & \multirow{2.5}{*}{\tabincell{c}{$L_1$ dist.\\to ORI}}\\
\cmidrule(lr){1-3} \cmidrule(l){4-5}
$L_{es}$&$L_{tf}$&$L_{ci}$& ResNet50 & DenseNet201 \\
\midrule
\ding{52}&\ding{52}&\ding{52} 		& $20.42\pm1.19$ & $28.11\pm1.76$ &$3.676$& $2.004$ \\
&\ding{52}&\ding{52}						& $37.45\pm1.19$ & $42.65\pm2.26$ & $3.695$ & $2.030$ \\ 
\ding{52}&&\ding{52} 						& $10.89\pm1.11$ & $11.71\pm0.85$ & $5.291$ & $17.724$ \\ 
\ding{52}&\ding{52}& 						& $11.02\pm0.43$ & $10.42\pm0.26$ & $4.610$ & $2.045$ \\ 
\bottomrule[1.2pt]
\end{tabular}
\begin{tablenotes}
\footnotesize
	\item[1] Comparisons are conducted on KCMI-550 in the classification task.
\end{tablenotes}
\end{threeparttable}
\end{center}
\vspace{-0.5cm}
\end{table}

\begin{figure}
\begin{center}
\begin{minipage}{0.326\linewidth}
  \centerline{\includegraphics[width=1\linewidth]{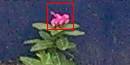}}
\end{minipage}
\hfill
\begin{minipage}{0.326\linewidth}
  \centerline{\includegraphics[width=1\linewidth]{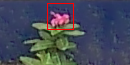}}
\end{minipage}
\hfill
\begin{minipage}{0.326\linewidth}
  \centerline{\includegraphics[width=1\linewidth]{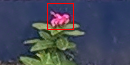}}
\end{minipage}
\vfill
\vspace{0.06cm}
\begin{minipage}{0.326\linewidth}
  \centerline{\includegraphics[width=1\linewidth]{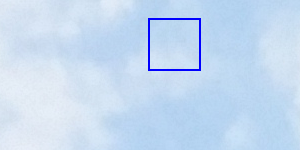}}
\end{minipage}
\hfill
\begin{minipage}{0.326\linewidth}
  \centerline{\includegraphics[width=1\linewidth]{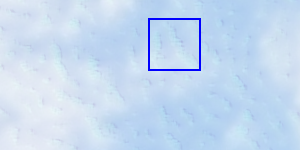}}
\end{minipage}
\hfill
\begin{minipage}{0.326\linewidth}
  \centerline{\includegraphics[width=1\linewidth]{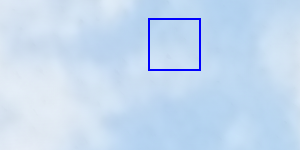}}
\end{minipage}
\vfill
\vspace{0.06cm}
\begin{minipage}{0.326\linewidth}
  \begin{minipage}{0.495\linewidth}
    \centerline{\includegraphics[width=1\linewidth]{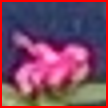}}
  \end{minipage}
  \hspace{-0.15cm}
  \begin{minipage}{0.495\linewidth}
    \centerline{\includegraphics[width=1\linewidth]{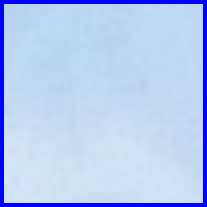}}
  \end{minipage}
\end{minipage}
\hfill
\begin{minipage}{0.326\linewidth}
  \begin{minipage}{0.495\linewidth}
    \centerline{\includegraphics[width=1\linewidth]{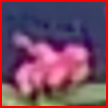}}
  \end{minipage}
  \hspace{-0.15cm}
  \begin{minipage}{0.495\linewidth}
    \centerline{\includegraphics[width=1\linewidth]{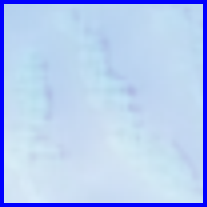}}
  \end{minipage}
\end{minipage}
\hfill
\begin{minipage}{0.326\linewidth}
  \begin{minipage}{0.495\linewidth}
    \centerline{\includegraphics[width=1\linewidth]{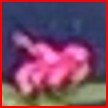}}
  \end{minipage}
  \hspace{-0.15cm}
  \begin{minipage}{0.495\linewidth}
    \centerline{\includegraphics[width=1\linewidth]{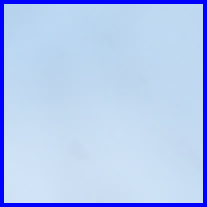}}
  \end{minipage}
\end{minipage}
\vfill
\vspace{0.1cm}
\begin{minipage}{0.326\linewidth}
  \centerline{(a) Original}
\end{minipage}
\hfill
\begin{minipage}{0.326\linewidth}
  \centerline{(b) Ours w/o $L_{ci}$}
\end{minipage}
\hfill
\begin{minipage}{0.326\linewidth}
  \centerline{(c) Ours}
\end{minipage}
\vfill
\end{center}
\vspace{-0.1cm}
\caption{Visual comparison for ablation study. Two image patches from KCMI-550 are adopted for comparison.}
\label{fig:ablation}
\vspace{-0.25cm}
\end{figure}

\begin{figure*}
\begin{center}
\begin{minipage}{0.02\textwidth}
\begin{sideways}
\fontsize{8}{9.6}{$im$ patch}
\end{sideways}
\end{minipage}
\hfill
\hspace{-0.1cm}
\begin{minipage}{0.094\linewidth}
  \centerline{\includegraphics[width=1\linewidth]{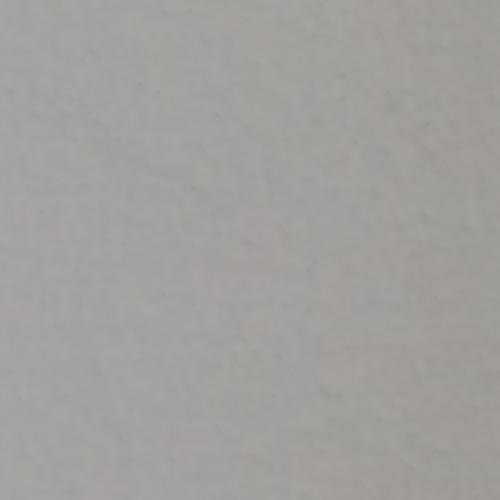}}
\end{minipage}
\hfill
\hspace{-0.1cm}
\begin{minipage}{0.094\linewidth}
  \centerline{\includegraphics[width=1\linewidth]{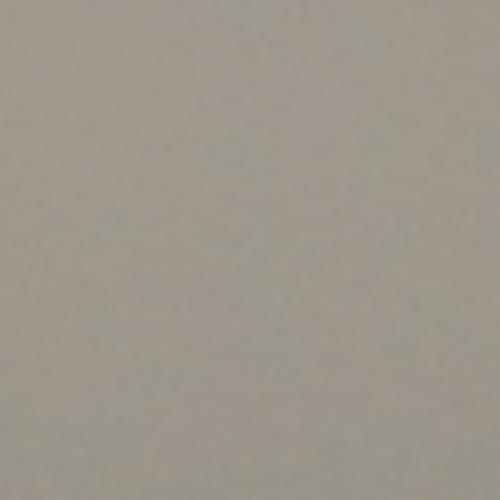}}
\end{minipage}
\hfill
\hspace{-0.1cm}
\begin{minipage}{0.094\linewidth}
  \centerline{\includegraphics[width=1\linewidth]{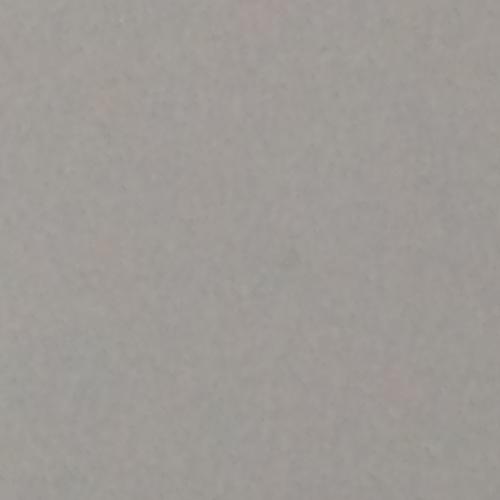}}
\end{minipage}
\hfill
\hspace{-0.1cm}
\begin{minipage}{0.094\linewidth}
  \centerline{\includegraphics[width=1\linewidth]{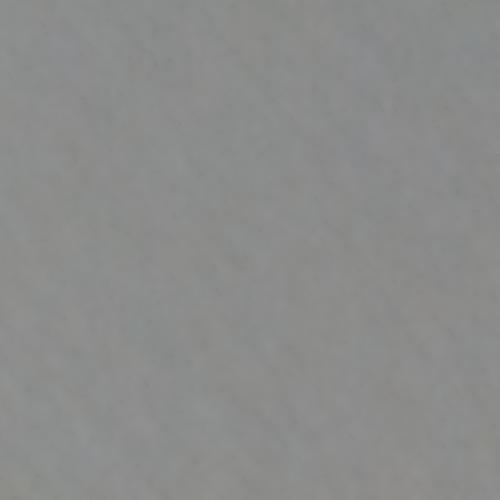}}
\end{minipage}
\hfill
\hspace{-0.1cm}
\begin{minipage}{0.094\linewidth}
  \centerline{\includegraphics[width=1\linewidth]{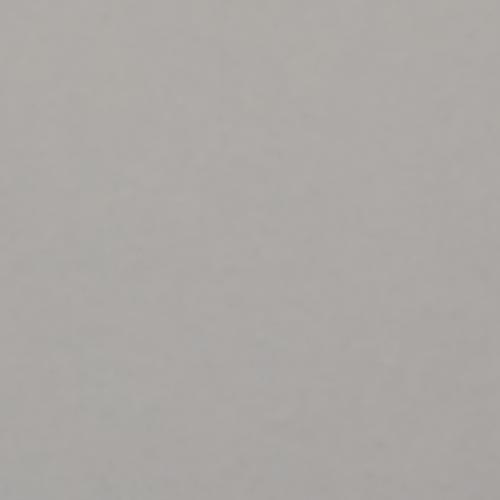}}
\end{minipage}
\hfill
\hspace{-0.1cm}
\begin{minipage}{0.094\linewidth}
  \centerline{\includegraphics[width=1\linewidth]{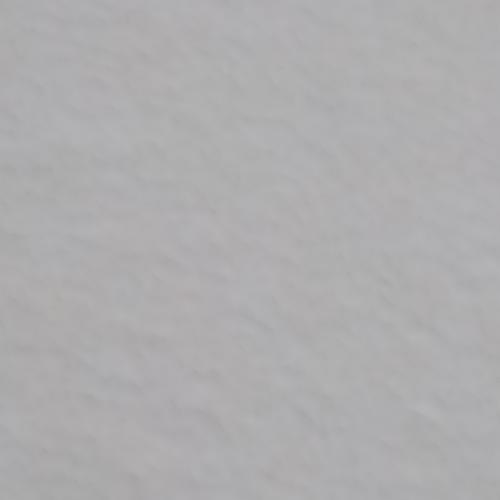}}
\end{minipage}
\hfill
\hspace{-0.1cm}
\begin{minipage}{0.094\linewidth}
  \centerline{\includegraphics[width=1\linewidth]{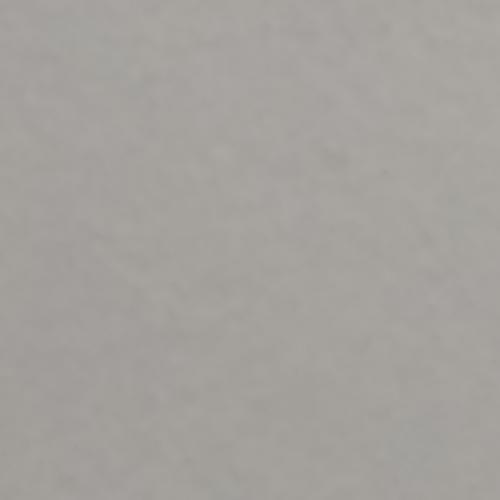}}
\end{minipage}
\hfill
\hspace{-0.1cm}
\begin{minipage}{0.094\linewidth}
  \centerline{\includegraphics[width=1\linewidth]{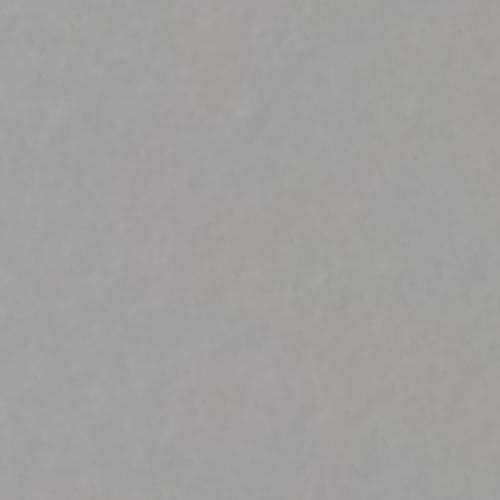}}
\end{minipage}
\hfill
\hspace{-0.1cm}
\begin{minipage}{0.094\linewidth}
  \centerline{\includegraphics[width=1\linewidth]{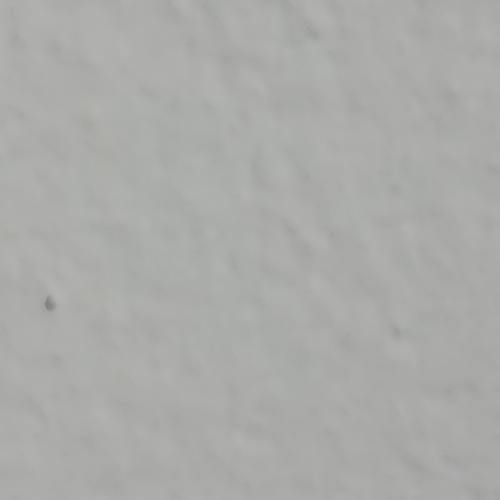}}
\end{minipage}
\hfill
\hspace{-0.1cm}
\begin{minipage}{0.094\linewidth}
  \centerline{\includegraphics[width=1\linewidth]{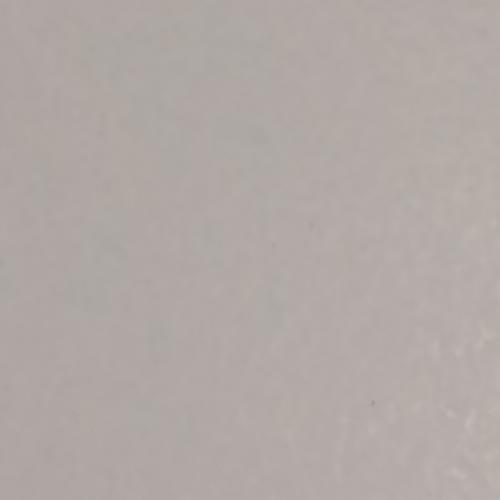}}
\end{minipage}
\vfill
\vspace{0.035cm}
\begin{minipage}{0.02\textwidth}
\begin{sideways}
\fontsize{8}{9.6}{$trs$ (spatial)}
\end{sideways}
\end{minipage}
\hfill
\hspace{-0.1cm}
\begin{minipage}{0.094\linewidth}
  \centerline{\includegraphics[width=1\linewidth]{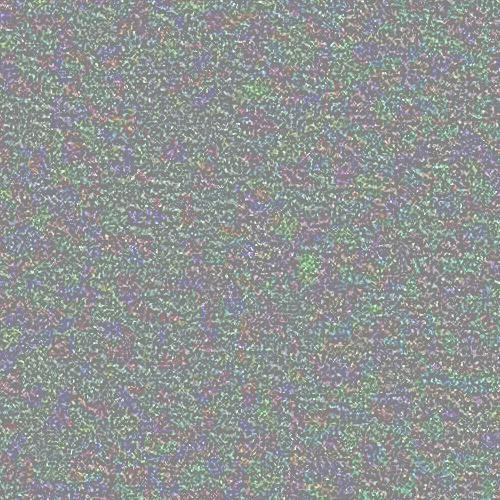}}
\end{minipage}
\hfill
\hspace{-0.1cm}
\begin{minipage}{0.094\linewidth}
  \centerline{\includegraphics[width=1\linewidth]{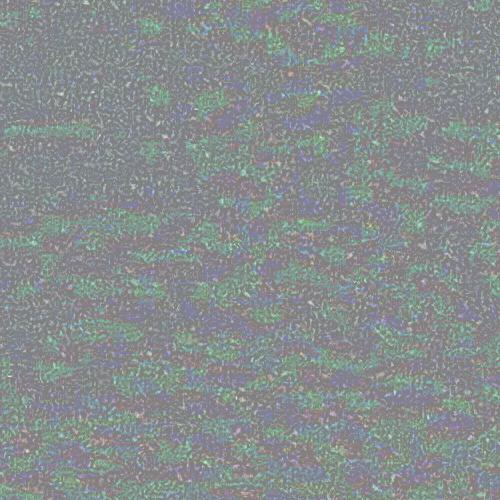}}
\end{minipage}
\hfill
\hspace{-0.1cm}
\begin{minipage}{0.094\linewidth}
  \centerline{\includegraphics[width=1\linewidth]{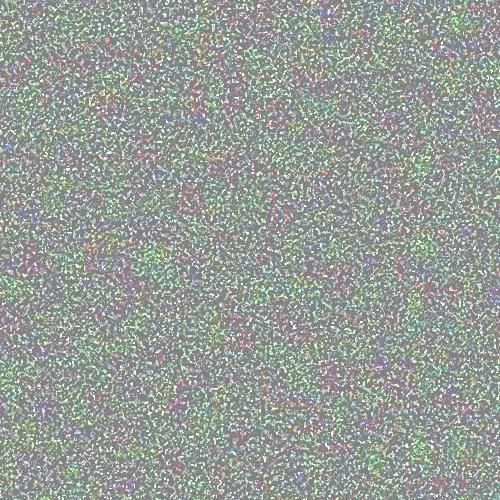}}
\end{minipage}
\hfill
\hspace{-0.1cm}
\begin{minipage}{0.094\linewidth}
  \centerline{\includegraphics[width=1\linewidth]{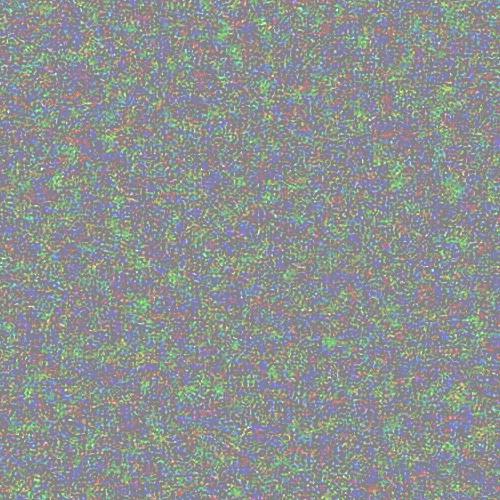}}
\end{minipage}
\hfill
\hspace{-0.1cm}
\begin{minipage}{0.094\linewidth}
  \centerline{\includegraphics[width=1\linewidth]{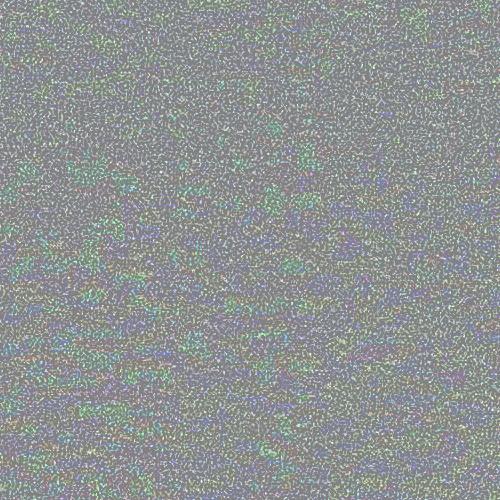}}
\end{minipage}
\hfill
\hspace{-0.1cm}
\begin{minipage}{0.094\linewidth}
  \centerline{\includegraphics[width=1\linewidth]{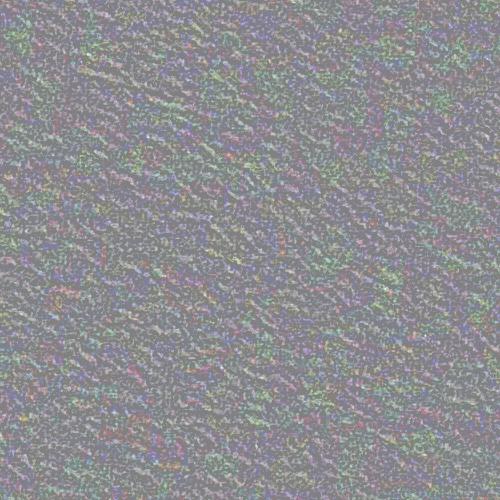}}
\end{minipage}
\hfill
\hspace{-0.1cm}
\begin{minipage}{0.094\linewidth}
  \centerline{\includegraphics[width=1\linewidth]{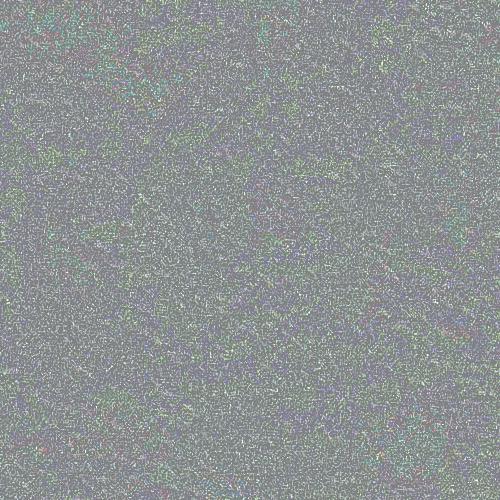}}
\end{minipage}
\hfill
\hspace{-0.1cm}
\begin{minipage}{0.094\linewidth}
  \centerline{\includegraphics[width=1\linewidth]{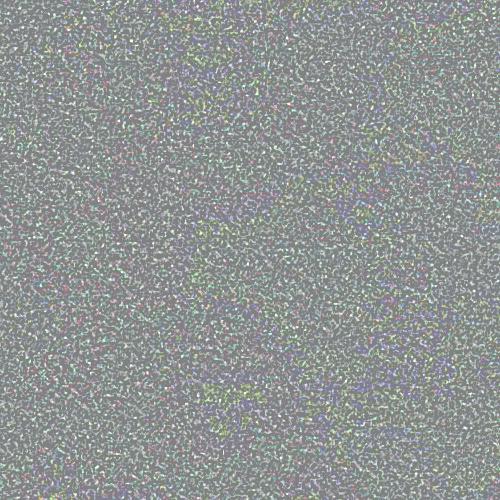}}
\end{minipage}
\hfill
\hspace{-0.1cm}
\begin{minipage}{0.094\linewidth}
  \centerline{\includegraphics[width=1\linewidth]{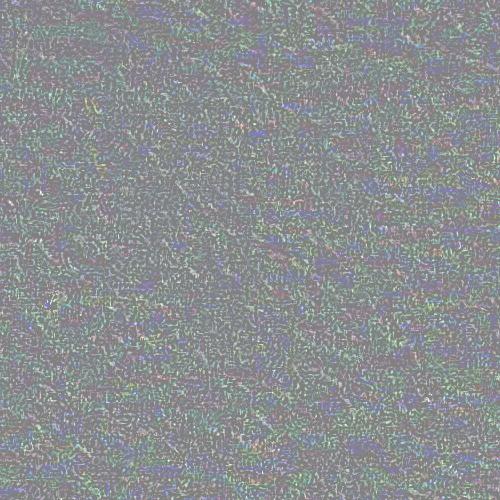}}
\end{minipage}
\hfill
\hspace{-0.1cm}
\begin{minipage}{0.094\linewidth}
  \centerline{\includegraphics[width=1\linewidth]{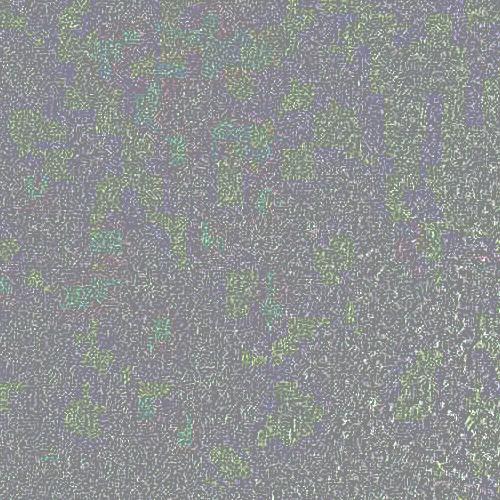}}
\end{minipage}
\vfill
\vspace{0.035cm}
\begin{minipage}{0.02\textwidth}
\begin{sideways}
\fontsize{8}{9.6}{$trs$ (freq.)}
\end{sideways}
\end{minipage}
\hfill
\hspace{-0.1cm}
\begin{minipage}{0.094\linewidth}
  \centerline{\includegraphics[width=1\linewidth]{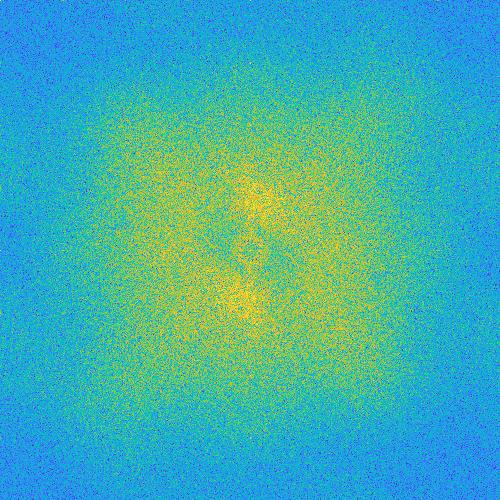}}
\end{minipage}
\hfill
\hspace{-0.1cm}
\begin{minipage}{0.094\linewidth}
  \centerline{\includegraphics[width=1\linewidth]{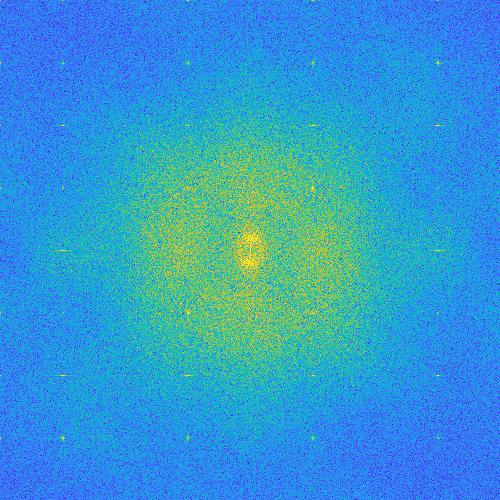}}
\end{minipage}
\hfill
\hspace{-0.1cm}
\begin{minipage}{0.094\linewidth}
  \centerline{\includegraphics[width=1\linewidth]{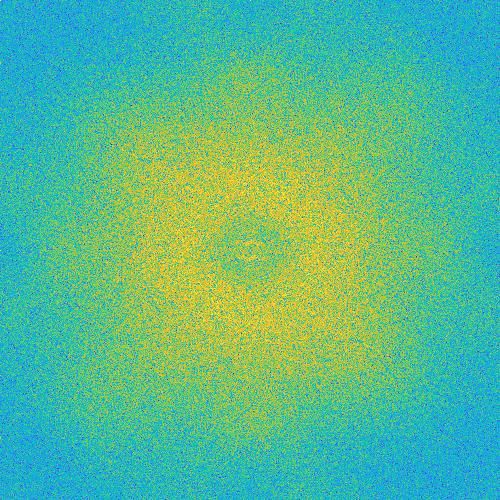}}
\end{minipage}
\hfill
\hspace{-0.1cm}
\begin{minipage}{0.094\linewidth}
  \centerline{\includegraphics[width=1\linewidth]{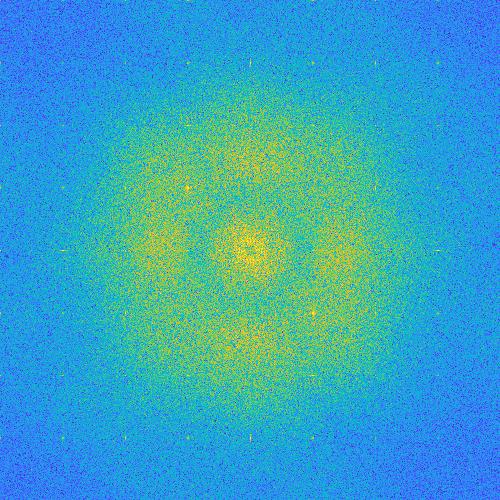}}
\end{minipage}
\hfill
\hspace{-0.1cm}
\begin{minipage}{0.094\linewidth}
  \centerline{\includegraphics[width=1\linewidth]{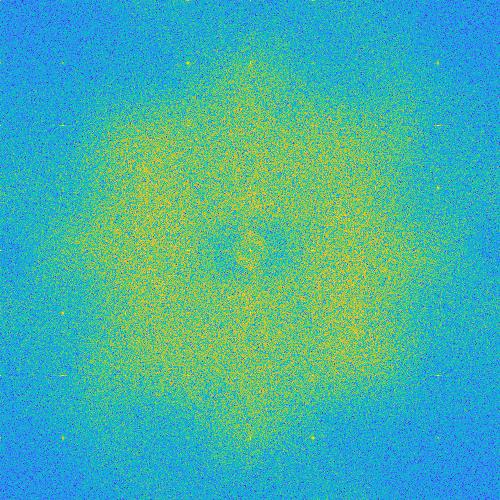}}
\end{minipage}
\hfill
\hspace{-0.1cm}
\begin{minipage}{0.094\linewidth}
  \centerline{\includegraphics[width=1\linewidth]{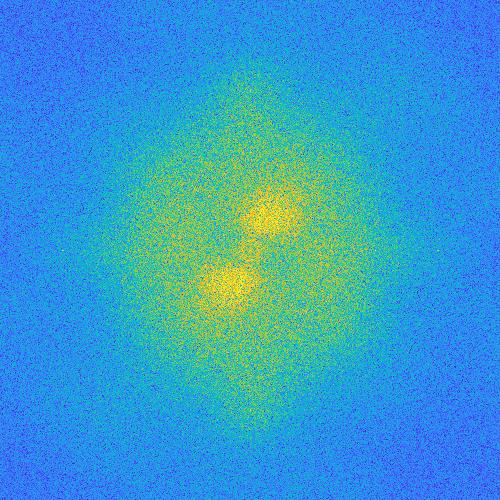}}
\end{minipage}
\hfill
\hspace{-0.1cm}
\begin{minipage}{0.094\linewidth}
  \centerline{\includegraphics[width=1\linewidth]{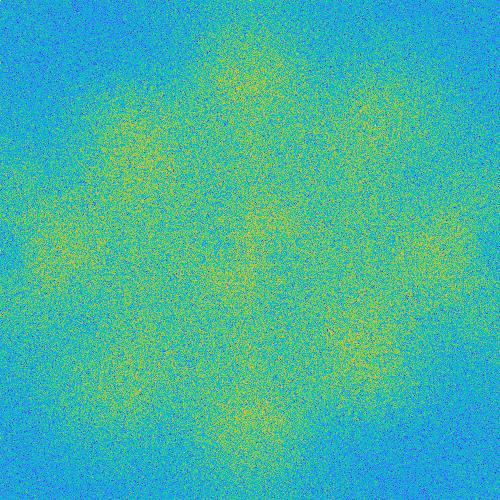}}
\end{minipage}
\hfill
\hspace{-0.1cm}
\begin{minipage}{0.094\linewidth}
  \centerline{\includegraphics[width=1\linewidth]{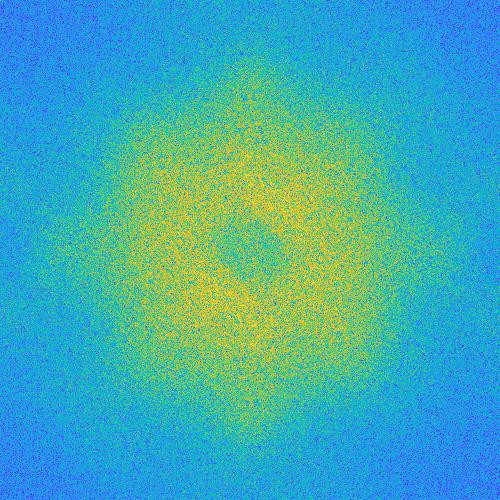}}
\end{minipage}
\hfill
\hspace{-0.1cm}
\begin{minipage}{0.094\linewidth}
  \centerline{\includegraphics[width=1\linewidth]{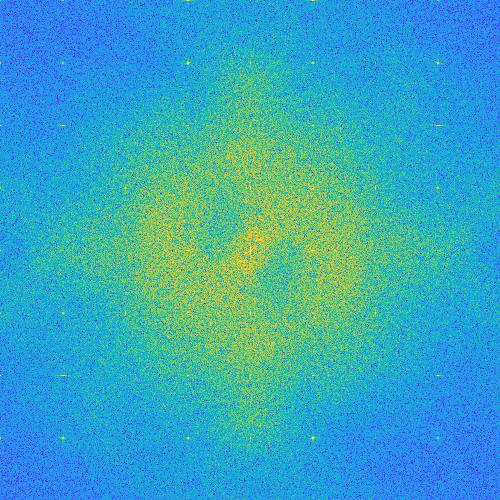}}
\end{minipage}
\hfill
\hspace{-0.1cm}
\begin{minipage}{0.094\linewidth}
  \centerline{\includegraphics[width=1\linewidth]{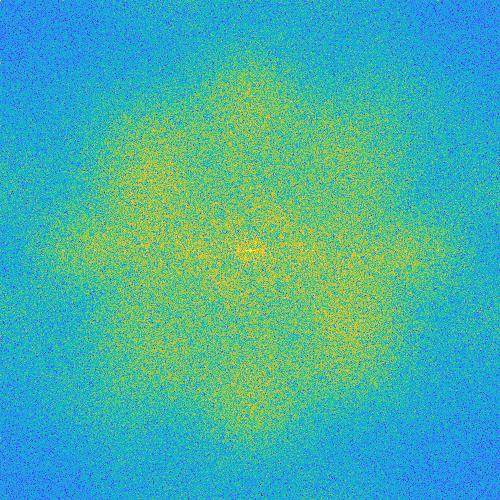}}
\end{minipage}
\vfill
\vspace{0.05cm}
\begin{minipage}{0.02\textwidth}
\begin{sideways}
\ 
\end{sideways}
\end{minipage}
\hfill
\hspace{-0.1cm}
\begin{minipage}{0.094\linewidth}
\fontsize{8}{9.6}\selectfont
  \centerline{iPhone 4S}
\end{minipage}
\hfill
\hspace{-0.1cm}
\begin{minipage}{0.094\linewidth}
\fontsize{8}{9.6}\selectfont
  \centerline{EVA-L09}
\end{minipage}
\hfill
\hspace{-0.1cm}
\begin{minipage}{0.094\linewidth}
\fontsize{8}{9.6}\selectfont
  \centerline{iPhone 5C}
\end{minipage}
\hfill
\hspace{-0.1cm}
\begin{minipage}{0.094\linewidth}
\fontsize{8}{9.6}\selectfont
  \centerline{Lenovo P70A}
\end{minipage}
\hfill
\hspace{-0.1cm}
\begin{minipage}{0.094\linewidth}
\fontsize{8}{9.6}\selectfont
  \centerline{VNS-L31}
\end{minipage}
\hfill
\hspace{-0.1cm}
\begin{minipage}{0.094\linewidth}
\fontsize{8}{9.6}\selectfont
  \centerline{RIDGE 4G}
\end{minipage}
\hfill
\hspace{-0.1cm}
\begin{minipage}{0.094\linewidth}
\fontsize{8}{9.6}\selectfont
  \centerline{Redmi Note3}
\end{minipage}
\hfill
\hspace{-0.1cm}
\begin{minipage}{0.094\linewidth}
\fontsize{8}{9.6}\selectfont
  \centerline{SM-G900F}
\end{minipage}
\hfill
\hspace{-0.1cm}
\begin{minipage}{0.094\linewidth}
\fontsize{8}{9.6}\selectfont
  \centerline{NEM-L51}
\end{minipage}
\hfill
\hspace{-0.1cm}
\begin{minipage}{0.094\linewidth}
\fontsize{8}{9.6}\selectfont
  \centerline{OnePlus A3003}
\end{minipage}
\vspace{0.15cm}
\caption{Visualization of camera trace extracted by SiamTE in spatial and frequency domains.
Image patches in smooth areas are cropped from VISION with a size of $500\times500$.
Brightness and contrast of camera trace in spatial domain are adjusted for a better visual experience.}
\label{fig:trace_visual}
\end{center}
\vspace{-0.5cm}
\end{figure*}

The gradient-based adversarial method is specially designed for CNN-based classifiers \cite{FGSM}. It thus achieves a satisfactory anti-forensic performance on CNN-based classifiers, which is on par with our proposed method (as listed in Table~\ref{tab:classification}).
However, the generalization ability of this adversarial method is poor. On one hand, the adversarial dithering is less effective than ours in the verification task, since a classic forensic method using hand-crafted features rather than a CNN-based one is adopted in this task (as listed in Table~\ref{tab:identification}).
On the other hand, the calculation of gradient is dependent on labels. Therefore, the adversarial method cannot generalize to images with unseen camera types, which makes it \textit{not} capable of processing images in the clustering task, thus not listed in Table~\ref{tab:clustering}.

Compared with the above baseline methods, our proposed SiamTE is more efficient for camera trace erasing. Specifically, at the similar (or lower) degree of manipulation, SiamTE significantly reduces the classification accuracy of ResNet50 from $99.80$\% to $20.42$\%, as listed in Table~\ref{tab:classification}.
In the clustering task, the performance of K-means clustering is halved by SiamTE, as listed in Table~\ref{tab:clustering}. In the verification task, SiamTE achieves $87.2$\% decrease on the correlation between an image and its camera type, as listed in Table~\ref{tab:identification}.
From the visual comparisons conducted in Figs.~\ref{fig:vcomp1} and \ref{fig:vcomp2}, the perceptual quality of our results are satisfactory, which is also verified by the lower NIQE values listed in Tables~\ref{tab:classification} and \ref{tab:clustering}.
The comprehensive experiments demonstrate a clear advantage of SiamTE for camera trace erasing over existing anti-forensic methods.

\subsection{Ablation study of hybrid loss}

Our proposed hybrid loss consists of three parts: (a) embedded similarity $L_{es}$, truncated fidelity $L_{tf}$, and cross identity $L_{ci}$. In this section, we provide ablation study to demonstrate the function of each part.
As listed in Table~\ref{tab:ablation}, each part of the hybrid loss has a contribution to the overall performance.
Without $L_{es}$, the anti-forensic performance in terms of classification accuracy is significantly weakened.
Without $L_{tf}$, the degree of manipulation loses control, which results in heavy destruction of content signal, as reflected by the large NIQE and $L_1$ distance.
Without $L_{ci}$, unfavorable artifacts are introduced to the visual results, which degrade the image quality and may reveal the adversarial process, as shown in Fig.~\ref{fig:ablation}.

\subsection{Analysis on camera trace}

\begin{figure}[t]
\begin{center}
\begin{minipage}{1.0\linewidth}
  \centerline{\includegraphics[width=0.99\linewidth]{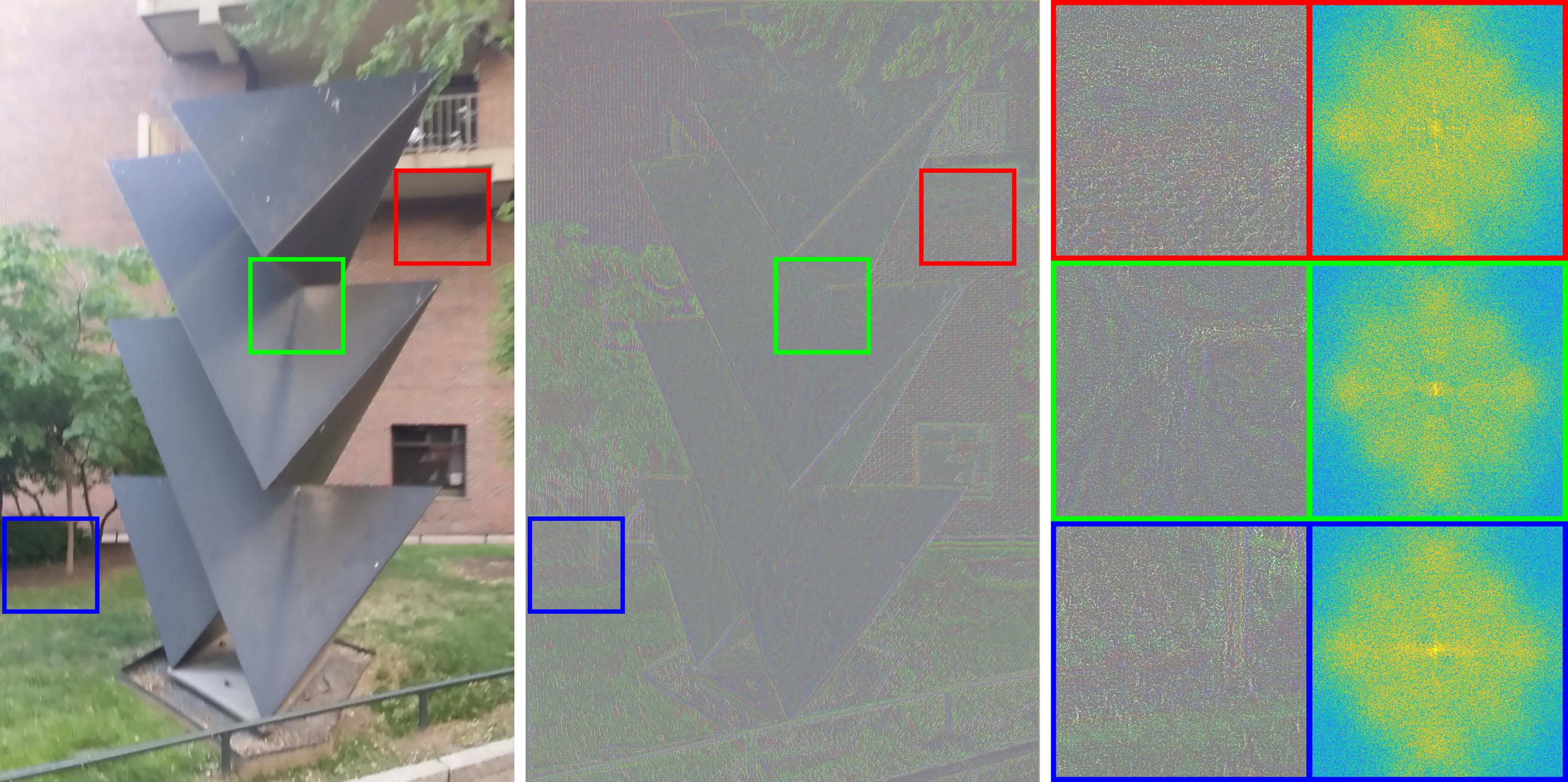}}
\end{minipage}
\end{center}
\vspace{-0.15cm}
\caption{Visualization of camera trace in a single image captured by GalaxyN3. From left to right: original image, camera trace extracted by SiamTE, and patches in spatial and frequency domains.}
\label{fig:trace_visual_im}
\vspace{-0.20cm}
\end{figure}

In this section, we analyze the extracted camera trace separately.
With a trained SiamTE, we extract camera trace from KCMI+ and KCMI-550, respectively.
We then train a DenseNet201 classifier on KCMI+ to identify origin from the extracted camera trace instead of the image itself.
On KCMI-550, we achieve an accuracy of $93.21\pm0.31$\% with camera trace, which is close to that of $99.87\pm0.12$\% with original images. It verifies that the extracted camera trace contains most of the camera-distinguishable information from original images.
We visualize the extracted camera trace in Fig.~\ref{fig:trace_visual}, which demonstrates that camera trace varies with different types of cameras. In comparison, patches in a single image have similar camera trace, as shown in Fig.~\ref{fig:trace_visual_im}.

\vspace{-0.05cm}
\section{Conclusion}
\vspace{-0.05cm}
We address a new low-level vision problem, termed as camera trace erasing, to reveal the weakness of trace-based forensic methods. It is of great importance to verify the security of image forensics.
We conduct a comprehensive investigation to existing anti-forensic methods, and propose SiamTE as an advanced solution, which significantly boosts the anti-forensic performance in three representative tasks.
We design a novel hybrid loss on the basis of Siamese architecture, which guides SiamTE to effectively erase camera trace without visible destruction of content signal.

\vspace{-0.1cm}
\section*{Acknowledgement}
\vspace{-0.05cm}
We acknowledge funding from National Key R\&D Program of China under Grant 2017YFA0700800, National Natural Science Foundation of China under Grant 61671419, and China Scholarship Council under Grant 201906340108.

{\small
\bibliographystyle{ieee}
\bibliography{camte}
}

\end{document}